%% file: template.tex
\documentclass[journal,compsoc]{IEEEtran}

\usepackage{multicol}
\usepackage{multirow}
\usepackage{pdflscape}
\usepackage{tabularx}
\usepackage[table]{xcolor}

\usepackage{tabu}                      %
\usepackage{booktabs}                  %
\usepackage{lipsum}                    %
\usepackage{mwe}                       %
\usepackage{microtype}
\usepackage{multirow}
\usepackage{longtable}  
\usepackage[numbers,square,comma,sort]{natbib}
\usepackage{booktabs}
\usepackage{tabularx}
\usepackage{multirow}
\usepackage[colorlinks=true,
            linkcolor=black,
            citecolor=blue,
            urlcolor=blue]{hyperref}

\usepackage{makecell}

\usepackage{array}
\renewcommand{\arraystretch}{1.3}
\newcolumntype{M}[1]{>{\centering\arraybackslash}m{#1}}
\newcolumntype{L}[1]{>{\raggedright\arraybackslash}m{#1}}

\usepackage{mathptmx}                  %
\usepackage{balance}
\begin{document}
\bstctlcite{IEEEexample:BSTcontrol}
\graphicspath{{figures/}{pictures/}{images/}{./}} %

\title{Augmented Reality User Interfaces for First Responders: A Scoping Literature Review}
\author{Erin Argo
\IEEEmembership{Student Member, IEEE}, 
Tanim Ahmed, 
Sarah Gable, 
Callie Hampton, \\
Jeronimo Grandi \IEEEmembership{Member, IEEE},
Regis Kopper, \IEEEmembership{Member, IEEE}
\thanks{E. Argo and J. Grandi are with the Department of Computer \& Cyber Sciences, Augusta University, Augusta, Georgia, 
USA}
\thanks{S. Gable and C. Hampton are with the Department of Computer Science, University of North Carolina at Greensboro,
Greensboro, North Carolina, USA.}
\thanks{Tanim Ahmed and R. Kopper is with the Department of Computer Science, Iowa State University,
Ames, Iowa, USA.}
}

\markboth{IEEE Transactioins on Visualization and Computer Graphics,~Vol.~XX, No.~XX, Month~20XX}%
{Augmented Reality Interfaces for First Responders}

\IEEEtitleabstractindextext{
  \begin{abstract}
  During the past decade, there has been a significant increase in research focused on integrating AR User Interfaces into public safety applications, particularly for first responders in the domains of Emergency Medical Services, Firefighting, and Law Enforcement. This paper presents the results of a scoping review involving the application of AR user interfaces in the public safety domain and applies an established systematic review methodology to provide a comprehensive analysis of the current research landscape, identifying key trends, challenges, and gaps in the literature. This review includes peer-reviewed publications indexed by the major scientific databases up to April 2025. A basic keyword search retrieved 1,751 papers, of which 90 were deemed relevant for this review. An in-depth analysis of the literature allowed the development of a faceted taxonomy that categorizes AR user interfaces for public safety. This classification lays a solid foundation for future research, while also highlighting key design considerations, challenges, and gaps in the literature. This review serves as a valuable resource for researchers and developers, offering insights that can drive further advances in the field.
  \end{abstract}

}

\maketitle

\input{Sections/1introduction}

\input{Sections/2methods_resources}

\input{Sections/3.0results}

\input{Sections/3.1paper_statistics}

\input{Sections/3.2taxonomy}

\input{Sections/3.3analysis}

\input{Sections/4discussion}

\input{Sections/5conclusion}

\balance

\section{Acknowledgments}
\noindent{This work was performed under award \#70NANB22H096 from the U.S. Department of Commerce, National Institute of Standards and Technology, Public Safety Communications Research Division. The authors acknowledge the use of AI-based language tools, specifically OpenAI’s ChatGPT 4o, to assist with proofreading, grammar revision, and stylistic editing during the preparation of this manuscript. No sections of the paper include AI-generated figures, images, or code. All scientific content, analysis, and conclusions are the sole responsibility of the authors}

\bibliographystyle{IEEEtran}

\bibliography{references}

\end{document}

%% file: Sections/1introduction.tex
\maketitle
\section{Introduction}\label{sec:22}

First responders play a critical role in emergency scenarios, and their ability to effectively assess, navigate, and respond to complex and high-pressure environments directly influences the outcomes of life-threatening situations. Whether it is paramedics caring for victims of an accident or firefighters combating hazardous fires, these professionals must constantly adapt to rapidly evolving and often unpredictable conditions ~\cite{ijerph15030534}. The complexity of such scenarios demands not only extensive training and experience, but also the support of advanced technologies to improve tasks such as communication, pathfinding, situational awareness, and decision making under stress~\cite{Kedia_2022}. 

While traditional tools and technologies have supported situational awareness (SA) and decision-making for first responders, the growing integration of Augmented Reality (AR) into public safety applications represents a significant leap forward in operational capabilities. 
AR's potential to streamline cognitive processing and reduce the time to action makes it particularly relevant in high-stress, time-sensitive situations faced by first responders. However, despite its promise, the deployment of AR in emergency scenarios raises questions about interface design, usability, and effectiveness under operational constraints. 

This scoping review evaluates the current landscape of AR technologies designed for first responders, emphasizing interface design, interaction modalities, and deployment challenges in real-world scenarios. We followed a structured systematic review methodology developed by Kitchenham and Charters~\cite{kitchenham2007guidelines} and conducted a comprehensive search and selection process across major academic databases such as ACM Digital Library, IEEE Xplore, and Scopus. Using the PICOC framework, defined by Petticrew and Roberts~\cite{petticrew2008systematic}, relevant studies were identified, screened, and analyzed. In addition to synthesizing key findings, this review proposes a high-level taxonomy to categorize AR user interfaces (UIs) for first responders, offering a structured approach to understanding current implementations and design patterns. By identifying trends, challenges, and gaps in the literature, the review aims to guide future research and improve AR system design in public safety operations.

Under the scope of AR UIs in public safety operations, this review systematically presents the scope of existing research, assesses the current advancements in the field, and categorizes identified interfaces. Objectively, it seeks to:

\begin{enumerate}
    \item Report the State of the Art by providing a comprehensive overview of the current advancements and practices of AR UIs in public safety operations.

    \item Analyze and Categorize AR UIs through the examination and classification of identified interfaces to better understand their applications and implications to public safety.

    \item Identify Gaps in the Literature by highlighting areas where existing research on public safety AR UI is lacking or underdeveloped.

\end{enumerate}

The research questions addressed in this review are:

\begin{itemize}
    \item RQ 1: What AR UIs have been proposed and/or used in the context of public safety?

    \item RQ 2: What AR UI elements have been demonstrated as effective for use in the context of public safety?

    \item RQ 3: What guidelines for the design and implementation of AR UIs for use in the context of public safety can be derived from the literature?

    \item RQ 4: What are the AR UI patterns used in public safety relevant for the definition of a taxonomy of such patterns?
\end{itemize}

%% file: Sections/2methods_resources.tex
\begin{table*}[t]
  \small
  \setlength{\tabcolsep}{6pt}
  \renewcommand{\arraystretch}{1.10}

  \caption{Search terms used in the systematic review, based on the PICOC framework. Terms are grouped by conceptual categories under Population and Intervention elements. Within each element, terms were combined using OR; between categories, using AND. The search targeted keywords in titles, abstracts, or keyword fields. Wildcards (*) were used to capture word variations.}
  \label{table:search_terms}
  \centering
  \begin{tabularx}{\textwidth}{@{}l l X@{}}
    \toprule
    \textbf{PICOC Element} & \textbf{Conceptual Category} & \textbf{Search Terms (Including Synonyms and Wildcards)} \\
    \midrule
    
    Population & \makecell[l]{First Responder \\Roles} & 
    EMS, EMT, Emergency Medic* Services, Fire fight*, First Respond*, 
    Emergency Respond*, Law Enforcement, Police Force, Policemen, 
    Policeman, Public Safety \\

    Intervention & \makecell[l]{Immersive Display \\Technologies} & 
    AR, XR, HUD, HMD, Heads Up Display*, Heads-Up Display*, 
    Head-Mounted Display*, Head Mounted Display*, Headmounted Display*, 
    Headmount Display*, Head mount Display* \\

    Intervention & Interface Modalities & 
    Heads Up Interface*, Heads-Up Interface*, Head-Mounted Interface*, 
    Head Mounted Interface*, Headmounted Interface*, Headmount Interface*, 
    Head mount Interface* \\

    Intervention & \makecell[l]{Reality-Altering \\Technologies} & 
    Augmented Realit*, Mixed Realit*, Extended Realit*, Integrated Realit*, 
    Augmented Interface*, Mixed Interface*, Extended Interface*, Integrated Interface* \\

    Intervention & Wearable Technologies & 
    Wearable Tech*, Wearable Device*, Wearable Computer*, Wearable Interface* \\

    \bottomrule
  \end{tabularx}
\end{table*}

\section{Methods and Resources}
This section describes the methodology of this article and details the systematic process undertaken during the Literature Review.

\subsection{Scope of the Literature Review}
This work adopts a scoping literature review approach, as defined by Arksey and O’Malley~\cite{arksey2005scoping}, which aims to map the breadth and nature of existing research on a given topic rather than to systematically evaluate study quality or outcomes. Unlike a systematic review or meta-analysis, which typically applies rigorous methodological filters and quality assessments to aggregate findings~\cite{mays2001systematic}, a scoping review seeks to identify and synthesize the full range of relevant studies, including emerging or underexplored areas, to provide a comprehensive overview of the field,˜\cite{peters2015guidance}.

In this review, papers were not specifically assessed for quality beyond the baseline requirement that all publications come from peer-reviewed journals or archival conference proceedings, including short papers, extended abstracts, and workshop papers. The goal was to capture the full scope of public safety AR UIs that have been proposed or prototyped, regardless of the level of implementation maturity. Consequently, AR UIs presented at the concept or low-fidelity prototype stage--for example, those tested in VR simulation environments or described as part of exploratory studies--were included in the review.

A key inclusion criterion was that the reviewed papers describe systems featuring visual augmentation. Systems that also incorporated additional modalities, such as auditory or haptic augmentation, were included, provided that visual augmentation was present. This inclusive scope was intended to capture the diversity of design approaches across the AR UI landscape for first responders, from early-stage research ideas to more advanced implementations. By focusing on breadth rather than strict implementation rigor, this scoping review aims to provide a broad foundation for future research, helping to identify design patterns, gaps, and emerging trends across the literature.

\subsection{Databases}
To find articles that represent the state-of-the-art in the literature, this paper considered 4 popular and highly impactful engineering and computing databases: the Association for Computing Machinery (ACM) Digital Library, the Institute of Electrical and Electronics Engineers (IEEE) Xplore Database, and the ProQuest and Scopus scientific indexing databases.

\subsection{Search Strategy}

We followed the methods outlined by Kitchenham and Charters~\cite{kitchenham2007guidelines}, who adapted a methodology initially focused on systematic reviews for the social sciences~\cite{petticrew2008systematic} to software engineering. The method advises creating a structured search strategy that begins with a Population, Intervention, Comparison, Outcome, and Context (PICOC) framework. This review focuses on a population of \textit{First Responders} and intervention of \textit{Augmented Reality}, with primary keywords derived from these categories. We decided not to restrict our review to comparison, outcome, or context elements. To develop a comprehensive and inclusive search term, we identified many synonyms for each element of our scoping review: First Responders and Augmented Reality. For example, some synonyms used in our search string for First Responders include terms like Emergency Medical Services (EMS), Emergency Medical Technician (EMT), Firefighter, and Law Enforcement. Similarly, Augmented Reality-related synonyms include AR, XR, Head-Mounted Display, and HMD. These synonyms were incorporated to ensure comprehensive coverage of the relevant literature during the search process.

Table \ref{table:search_terms} shows all the terms utilized in the systematic review. Using these related terms, a base search string was formulated to guide the systematic search.

\begin{figure}
   \centering
   \includegraphics[width=\linewidth]{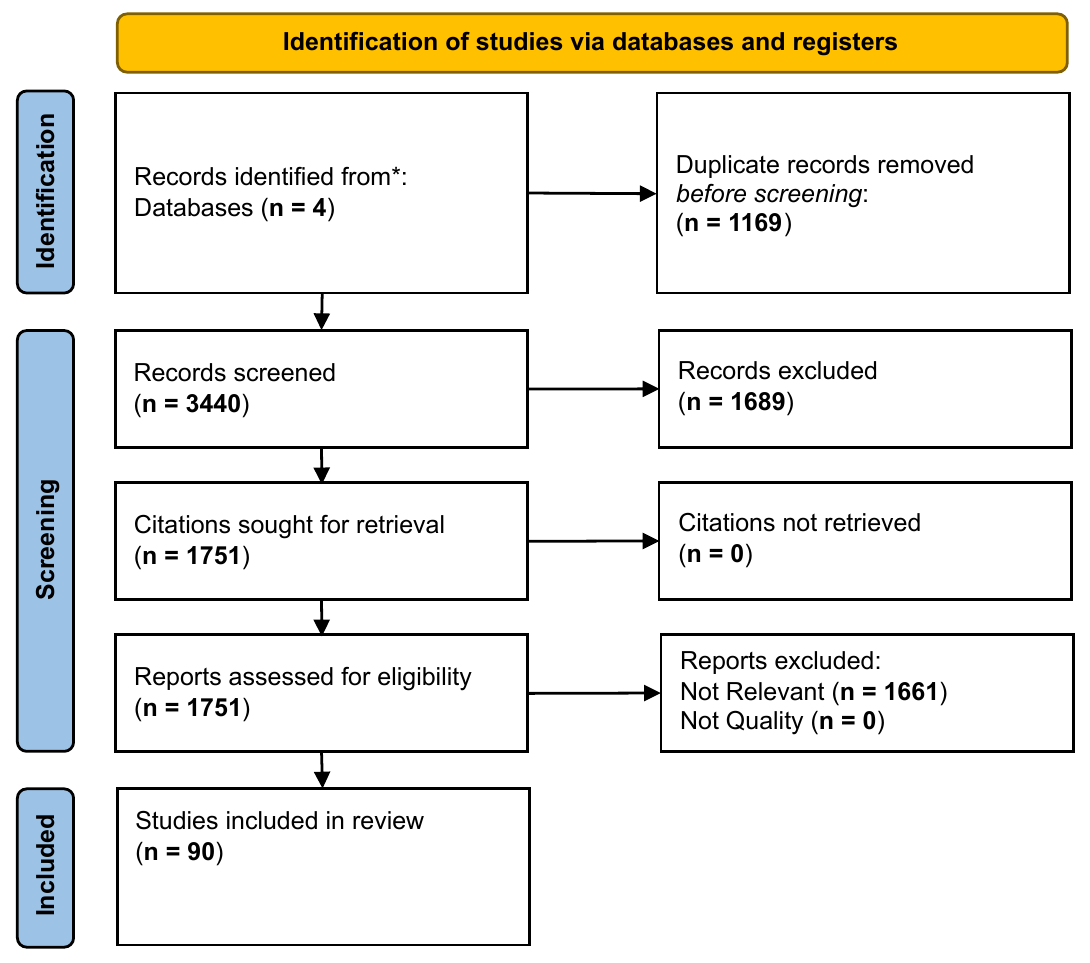}
   \caption{PRISMA 2020 flow diagram summarizing study selection for the scoping review. The process resulted in 90 studies that were included in the final synthesis.}
   \label{fig:screening_flowchart}
\end{figure}

This base string was adapted for the databases to optimize performance and to ensure that all key terms remained present. In each case, papers were automatically filtered to only include those where the keyword appeared in the title, abstract, or keywords.

Finally, certain databases provided tools by which papers constrained by subject matter and publication type.  The Scopus database allowed filtering by subject matter, and the search included articles in the following subject areas (categories): Health Sciences (Medicine, Nursing, Health Professions), Physical Sciences (Computer Science, Engineering, Mathematics, Physics and Astronomy), and Social Sciences (Decision Sciences, and Psychology, and Social Sciences). For Scopus, IEEE Xplore, and ACM Digital Library, we also constrained the search by the type of publication to only include peer-reviewed publications: Articles, Reviews, Conference Papers, and Short Surveys.

Figure \ref{fig:screening_flowchart} outlines the sequential screening and eligibility decisions applied to the records retrieved from all databases. It shows how the initial search results were merged and de-duplicated, how titles and abstracts were screened, how full-text articles were assessed for eligibility, and, finally, how the studies included in the review were selected.

\subsection{Selection Process}
A total of 1,751 papers resulted from the keyword search and their citations were retrieved. These citations were extracted and uploaded into an online tool called Parsifal~\cite{parsifal}, an online tool to support the cataloging of articles and data extraction for systematic reviews. Inclusion criteria we relevance, availability, and language (English). No articles were excluded due to language or lack of availability. The assessment for relevance included, initially, a review of the titles and abstracts of the papers. When title and abstract screening was not sufficient to determine relevance, the researcher skimmed the paper to make an informed decision. At least two members of the research team independently assessed the papers for relevance. Whenever a disagreement existed, the article was brought up for discussion among the senior members of the research team and a consensus decision was made.

Of the initial 1,751 retrieved papers, 1,661 were rejected as irrelevant to the scoping review. Thus, 90 papers were included in the scoping review of the literature.

\begin{figure}[t]
    \centering
    \includegraphics[width=\linewidth]{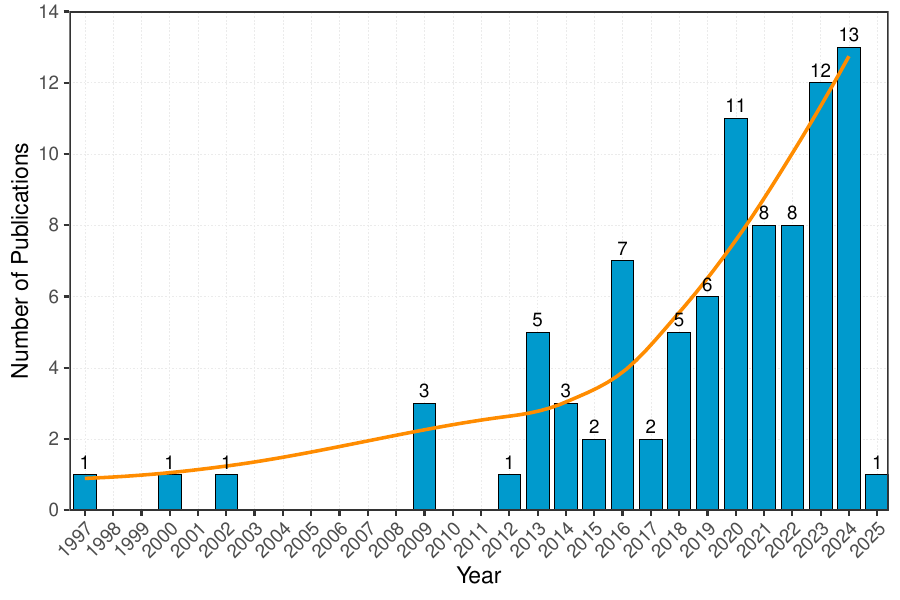}
    \caption{Annual research output on AR UIs for first responders. Publication activity was sporadic during the first decade. Output begins to climb after 2008, coinciding with the first mobile-AR toolkits and SDKs. A pronounced surge follows the arrival of consumer XR hardware: Oculus Rift DK1 in 2013, Microsoft HoloLens and SteamVR head-sets in 2016, Magic Leap One in 2018, and Meta Quest in 2019–2020. The single paper recorded for 2025 reflects our literature search cut-off at April 2025 and should not be interpreted as a decline.}
    \label{fig:pxy}
\end{figure}

\subsection{Data Extraction}
Each of the 90 papers included in this review was read in its entirety by two members of the research team. Upon reading the paper, the researcher filled out an initial data extraction form that included items such as type of research, research questions, software used in development, type of study (comparative or not), a summary of the paper, type of hardware used, and research challenges.

At a second stage, the same two researchers filled out a different form to extract more objective and quantitative data about the paper, including: year of conduct, country of origin, presence of formal evaluation, evaluation type, measures and independent variables used, questionnaires used, number of and details about participants, technology used, public safety discipline, and relevant classification criteria (requirements, type of AR UI, application).

\subsection{Method Limitations}
This scoping review presents certain limitations that should be acknowledged.  By the nature of keyword-based searches, potentially relevant articles that did not match the search string were not included in the review. This may have led some relevant studies to be excluded from the review. Additionally, some domains of knowledge have not been researched due to the search being limited to the chosen databases. Despite these limitations, we designed the search criteria to align closely with the research questions, target population (first responders), and intervention (AR UIs), maximizing inclusion. Moreover, we conducted the search within high-impact, comprehensive digital libraries and databases. As a result, we expect that any missed or excluded studies have minimal influence on the overall conclusions and analysis presented in this review.

%% file: Sections/3.0results.tex
\section{Results}
This section presents the findings of the scoping literature review. It begins with an overview of publication statistics, highlighting trends in research output over time and by geographical distribution. Subsequently, we introduce a faceted taxonomy developed from the analyzed literature to categorize AR user interfaces for first responders. The core of this section is a detailed analysis of the various AR interface elements identified in the reviewed papers, organized by their primary function and characteristics.

%% file: Sections/3.1paper_statistics.tex
\subsection{Publication Trends}

\begin{figure}[t]
    \centering
    \includegraphics[width=\linewidth]{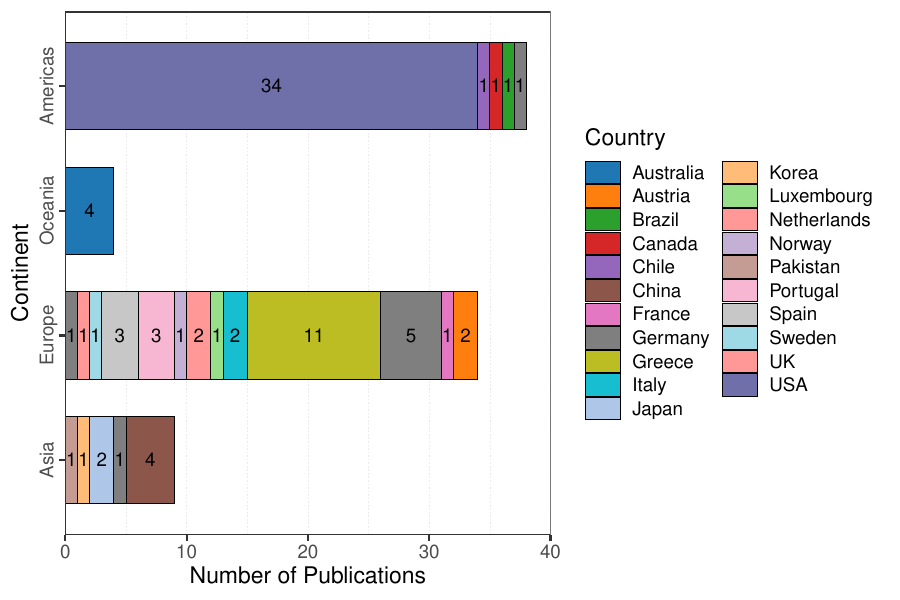}
    \caption{Geographic distribution of peer-reviewed papers on augmented-reality user interfaces for first responders. Horizontal bars give the continent-level totals, with each segment colour-coded by the country that authored the work.}
    \label{fig:pxc}
\end{figure}
    
Understanding the state of the art of AR 3D UIs within the context of public safety is essential for both researchers and first responders. To investigate this, this scoping review covers publications from the past 28 years, revealing trends in publication frequency and geographic distribution. As shown in Figure \ref{fig:pxy} the number of publications per year initially spiked in 2009, followed sustained increase after 2016. This suggests a growing research interest and an expanding body of work focused on AR UIs in public safety applications.

\begin{figure*}[ht!]
    \centering
    \includegraphics[width=\textwidth]{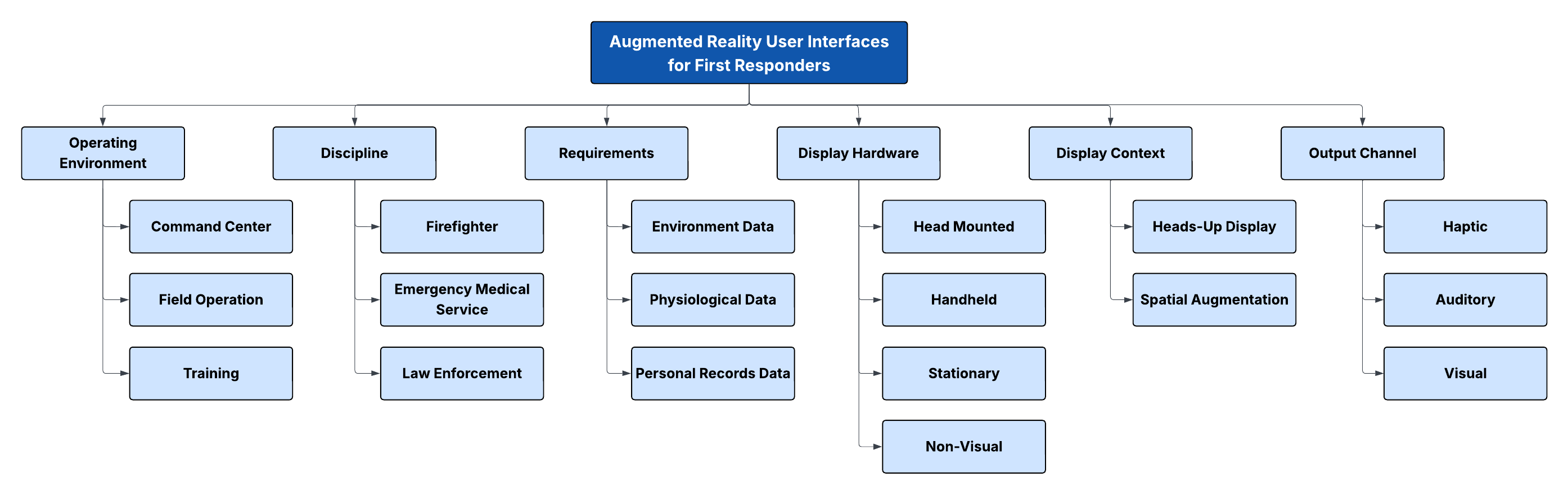}
    \caption{Taxonomy of AR UI designs for first responder applications developed as a result of the analysis of the literature covered in this scoping review. The diagram organizes the 90 studies into six orthogonal design facets. Any combination of facet values maps to a potential system configuration, for example, a head-mounted heads-up display that delivers visual and haptic alerts to a firefighter during field operations.}
    \label{fig:taxonomy}
\end{figure*}

Figure \ref{fig:pxc} presents a geographical distribution of papers by continent and country. The data show that research on AR UIs for first responders is not geographically concentrated; rather, contributions come from a wide range of country across Asia, Europe, Oceania, and the Americas. Notably, the United States accounts for a substantial portion of the publications, underscoring its prominent role in advancing this field. Overall, the global distribution of research efforts highlights the widespread recognition of AR's potential to enhance the safety and effectiveness of first responders by integrating advanced technology into their operational workflows.

%% file: Sections/3.2taxonomy.tex
\subsection{Taxonomy}

\input{Sections/Tables/operating_environment_table}
In analyzing the literature, this paper aims to identify trends and gaps in current research. To this end, we developed a high-level faceted taxonomy, shown in figure \ref{fig:taxonomy}, based on the current literature as a model for categorization of existing literature on AR UIs for first responders.

The six facets presented in Figure \ref{fig:taxonomy}--Operating Environment, Discipline, Requirements, Display Hardware, Display Context, and Output Channel--were selected because they consistently emerged from the reviewed literature as the most fundamental high-level distinctions characterizing AR systems for first responders. \textbf{Operating Environment} defines the operations context in which the AR UI is proposed. \textbf{Discipline} identifies the public safety service (EMS, Firefighting, Law Enforcement) for which the AR UI is applied. \textbf{Requirements} dictate the system's core informational function by specifying the primary data types the AR system is built on to process and present. \textbf{Display Hardware} identifies the type of AR technology that is used for the AR UI. \textbf{Display Context} classifies how the augmented information is rendered to the user, whether spatially or as a heads-up display (HUD).
Lastly, \textbf{Output Channel} covers the sensory modalities for information         
delivery. Collectively, these facets offer a robust framework for classifying the     
diverse AR applications in public safety and for understanding their core design          
attributes. 

This taxonomy unifies several higher-level concepts identified in the literature into structured facets. These facets were consistently represented across the reviewed papers, with few exceptions. Each paper was classified according to the facets. However, as a high-level classification, the goal of the taxonomy is not to describe specific interfaces in detail. For that purpose, Section \ref{sec:analysis} offers a detailed analysis of the literature according to the AR user interfaces that have been proposed in the identified literature. 

The following subsections outline the facets identified from the literature that influenced the overall purpose and features of the interfaces developed.

\subsubsection{Operating Environment}

This facet refers to the operational context for which a given interface is developed. \textbf{Field operation} refers to any interface used outside of training or command-and-control environments, involving live incidents and real-time dynamic environments. \textbf{Command center} refers to interfaces designed for incident commanders to coordinate their teams. A \textbf{training} operating environment is one where AR is employed to train first responders to perform public safety operations. In this case, AR may or may not be used in the actual operation itself, but it is specifically used during training. For example, if AR is used to help first responders learn a procedure by, for example, providing augmented overlays that will not be present in the real operational procedure, such an interface would be classified as training. Conversely, training interfaces that use virtual reality as a simulation platform for AR operations are not classified in the taxonomy, as they do not directly involve AR. Table \ref{table:environment} presents the reviewed publications classified by operating environment.

\subsubsection{Discipline}

\input{Sections/Tables/discipline_table}

This facet identifies the group of first responders targeted by the AR UI. It corresponds to the Population criterion in the PICOC framework~\cite{petticrew2008systematic}. The categories include \textbf{Law Enforcement}, \textbf{Emergency Medical Services}, and \textbf{Firefighting}. For classification purposes, the category \textbf{Others} is used when the target population did not fall in these three groups, such as the military or general public. In our scoping review, four papers targeted the ``Other'' category, meaning that the interface was proposed for public safety applications beyond the three main disciplines, such as Mantoro et al.~\cite{mantoro_pathfinding_2021}, who developed AR interfaces disaster evacuation. Additionally, three papers developed interfaces that supported crossover between two disciplines~\cite{grandi_design_2021,wilchek2025ajna,nalamothu2024leveraging} designed systems applicable to law enforcement and firefighting. Table \ref{table:discipline} presents the reviewed publications classified by discipline.

\subsubsection{Requirements}

\input{Sections/Tables/requirements_table}

This facet categorizes AR systems based on the primary sources of data they process or display to support first responder tasks. These \textit{Data} Requirements are fundamental to the system's purpose. \textbf{Environment Data} includes information sourced from the operational surroundings, such as geospatial data, object locations, or hazard information. \textbf{Physiological Data} covers real-time biological measurements from patients or first responders. \textbf{Patient/Suspect Data} encompasses individual-specific information, such as patient's medical history or suspect's records. The category \textbf{Context Unaware} refers to systems where the core AR functionalities do not on, or do not report, strong reliance on real-time data streams; such systems may, for example, provide generic procedural guidance or display static information. The methods of data acquisition (e.g., sensors, databases) are not the focus here--instead, the taxonomy focuses on the data source the AR interface is built to handle. Table \ref{table:requirements} classifies the reviewed publications according to their requirements.

\subsubsection{Display Hardware}

\input{Sections/Tables/display_hardware_table}

This faced refers to the type of AR hardware used for feedback and interface display. We identified three categories: \textbf{handheld devices}, such as mobile phones or smart watches, \textbf{head-mounted displays}, such as VST or OST HMDs, and \textbf{stationary devices}, including desktops or touch-walls. Table \ref{table:display_hardware} includes citations to the papers classified according to the display hardware they use.

\subsubsection{Display Context}

\input{Sections/Tables/display_context_table}

This facet describes how visual elements are integrated and displayed to the user. We identified two main contexts, based on whether the content is displayed specially within the environment or on the screen plane. \textbf{Spatial Augmentation} describes any projection of two- or three-dimensional virtual objects onto the real environment, independent of the display device's orientation. \textbf{Heads-Up Display (HUD)} classifies  two- or three- dimensional interfaces tied to the display device's orientation and displayed within the user's Field of View. Table \ref{table:display_context} shows the classification of the reviewed literature with respect to display context.

Section \ref{sec:analysis} breaks down spatial augmentations and heads-up displays into interface elements. Some of these elements traditionally considered HUD components can also be spatially augmented and vice-versa, characterizing dynamic interface implementations.

\subsubsection{Output Channel}

The output channel defines the mode of feedback displayed to the user, including \textbf{Visual}, \textbf{Auditory}, and \textbf{Haptic} channels. While it is possible to have purely haptic or auditory AR interfaces, this review only includes interfaces that feature visual augmentation, but may additionally contain other output channels. The classification of papers according to this facet is shown in Table~\ref{table:output}.

While this taxonomy provides a high-level categorization, the section~\ref{sec:analysis} analyzes the specific AR interface elements encountered in the literature, offering a more granular understanding of the AR UIs developed for first responders.

%% file: Sections/Tables/operating_environment_table.tex
\begin{table*}[t]
  \small
  \setlength{\tabcolsep}{6pt}
  \renewcommand{\arraystretch}{1.10} 

  \caption{Publications classified by operating environment}
  \label{table:environment}
  \centering
  \begin{tabularx}{\textwidth}{@{}l l c X@{}}
    \toprule
    \textbf{Facet} & \textbf{Category} & \textbf{Count} & \textbf{Publications} \\
    \midrule

      \makecell[l]{Operating\\Environment} & Field & 80 &
      \cite{antevski_5g-based_2021, azpiroz2024white,
            schmorrow_implementing_2016,
            berndt_optical_2015,
            bhattarai_embedded_2020,
            bram-larbi_collision_2020,
            brandao_using_2017,
            broach_usability_2018,
            cai_mobile_2023,
            campos_mobile_2019,
            chae_design_2023,
            chalimas_cross-device_2023,
            chan_erwear_2016,
            cheng_effective_2023,
            christaki_augmented_2022,
            christian_google_2014,
            chroust2009training,
            crowley_ar_2014,
            dave_augmenting_2013,
            davis2002augmented,
            demirkan_underground_2024,
            dianxi_design_2021,
            engelbrecht_viability_2018,
            del2014red,
            follmann2019technical,
            garcia_health-5g_2023,
            garcia2024smart,
            gkika_object_2023,
            grandi_design_2021,
            grandi_simulating_2020,
            arregui_augmented_2022,
            haque_augmented_2020,
            ii_mixed_2020,
            hu2021human,
            hu_seeing_2022,
            ierache_augmented_2016,
            ismael_radiological_2023,
            kapalo_comparing_2022,
            karakostas_real-time_2024,
            kelley_guiding_2024,
            kim_mobile_2016,
            lackey_new_2016,
            li2019fire,
            mannuru_mobile_2022,
            mantoro_pathfinding_2021,
            nalamothu2024leveraging,
            nelson_user-centered_2022,
            newaz_supporting_2015,
            schmorrow_mixed-methods_2024,
            nunes2019augmented,
            nunes_augmented_2018,
            oconnor_augmented_2023,
            oconnor_pilot_2024,
            oregui2024augmented,
            pavlopoulos_augmented_1997,
            phillips2020robotic,
            balfour_next_2012,
            rainer2009simrad,
            frasson_cpr_2023,
            rojas-munoz_evaluation_2020,
            safranoglou2024augmented,
            sainidis_single-handed_2021,
            sampson_ar_2022,
            schlosser_head-worn_displays_2021,
            schlosser2024effects,
            schonauer_3d_2013,
            sebillo_training_2016,
            sharma_situational_2020,
            siu_sidebars_2013,
            stefanidi_real-time_2022,
            tadokoro_robocup-rescue_2000,
            tiemann_celidon_2020,
            umlauft_communication_2016,
            walker_mixed_2021,
            wang_person--interest_2013,
            wang_method_2023,
            weichelt_augmented_2018,
            whitlock_designing_2019,
            wilchek2025ajna,
            zhang_exploring_2024}
      \\[4pt] 

      & Command Center & 6 &
      \cite{arif_comparative_2019,
            balfour_what_2013,
            lugtenberg_magicbook_2023,
            nilsson_using_2009,
            peretti_augmented_2022,
            stone_mixed_2017}
      \\[4pt]

      & Training & 3 &
      \cite{doswell2020juxtopia,
            friedman2024prehospital,
            koutitas2019virtual}
      \\[4pt]

      & Others & 1 &
      \cite{majumdar_cloud-based_2023}
      \\

    \bottomrule
  \end{tabularx}
  \normalsize
\end{table*}

%% file: Sections/Tables/discipline_table.tex
\begin{table*}[t]
  \small
  \setlength{\tabcolsep}{6pt}
  \renewcommand{\arraystretch}{1.10}

  \caption{Publications classified by targeted public safety discipline}
  \label{table:discipline}
  \centering
  \begin{tabularx}{\textwidth}{@{}l l c X@{}}
    \toprule
    \textbf{Facet} & \textbf{Category} & \textbf{Count} & \textbf{Publications} \\
    \midrule

      Discipline & EMS & 55 &
      \cite{antevski_5g-based_2021,
            balfour_what_2013,
            berndt_optical_2015,
            bram-larbi_collision_2020,
            broach_usability_2018,
            cai_mobile_2023,
            campos_mobile_2019,
            chan_erwear_2016,
            christaki_augmented_2022,
            chroust2009training,
            crowley_ar_2014,
            dave_augmenting_2013,
            davis2002augmented,
            demirkan_underground_2024,
            doswell2020juxtopia,
            del2014red,
            follmann2019technical,
            friedman2024prehospital,
            garcia_health-5g_2023,
            arregui_augmented_2022,
            ii_mixed_2020,
            hu2021human,
            hu_seeing_2022,
            ierache_augmented_2016,
            ismael_radiological_2023,
            kelley_guiding_2024,
            kim_mobile_2016,
            koutitas2019virtual,
            lackey_new_2016,
            nelson_user-centered_2022,
            newaz_supporting_2015,
            nilsson_using_2009,
            nunes2019augmented,
            nunes_augmented_2018,
            oconnor_augmented_2023,
            oconnor_pilot_2024,
            pavlopoulos_augmented_1997,
            peretti_augmented_2022,
            balfour_next_2012,
            rainer2009simrad,
            frasson_cpr_2023,
            rojas-munoz_evaluation_2020,
            safranoglou2024augmented,
            sainidis_single-handed_2021,
            schlosser_head-worn_displays_2021,
            schlosser2024effects,
            sebillo_training_2016,
            siu_sidebars_2013,
            stone_mixed_2017,
            tadokoro_robocup-rescue_2000,
            tiemann_celidon_2020,
            umlauft_communication_2016,
            walker_mixed_2021,
            wang_method_2023,
            zhang_exploring_2024}
      \\[4pt]

      & Firefighter & 53 &
      \cite{arif_comparative_2019,
            azpiroz2024white,
            schmorrow_implementing_2016,
            balfour_what_2013,
            bhattarai_embedded_2020,
            cai_mobile_2023,
            campos_mobile_2019,
            chae_design_2023,
            chalimas_cross-device_2023,
            chan_erwear_2016,
            cheng_effective_2023,
            christaki_augmented_2022,
            chroust2009training,
            crowley_ar_2014,
            dave_augmenting_2013,
            demirkan_underground_2024,
            dianxi_design_2021,
            garcia2024smart,
            gkika_object_2023,
            grandi_design_2021,
            arregui_augmented_2022,
            ii_mixed_2020,
            hu2021human,
            hu_seeing_2022,
            ismael_radiological_2023,
            kapalo_comparing_2022,
            kelley_guiding_2024,
            kim_mobile_2016,
            li2019fire,
            lugtenberg_magicbook_2023,
            majumdar_cloud-based_2023,
            nalamothu2024leveraging,
            newaz_supporting_2015,
            nilsson_using_2009,
            nunes2019augmented,
            nunes_augmented_2018,
            oregui2024augmented,
            peretti_augmented_2022,
            balfour_next_2012,
            rainer2009simrad,
            safranoglou2024augmented,
            sainidis_single-handed_2021,
            schonauer_3d_2013,
            sebillo_training_2016,
            siu_sidebars_2013,
            tadokoro_robocup-rescue_2000,
            tiemann_celidon_2020,
            umlauft_communication_2016,
            walker_mixed_2021,
            weichelt_augmented_2018,
            whitlock_designing_2019,
            wilchek2025ajna,
            zhang_exploring_2024}
      \\[4pt]

      & Law Enforcement & 45 &
      \cite{balfour_what_2013,
            brandao_using_2017,
            cai_mobile_2023,
            campos_mobile_2019,
            chan_erwear_2016,
            christaki_augmented_2022,
            christian_google_2014,
            chroust2009training,
            crowley_ar_2014,
            dave_augmenting_2013,
            demirkan_underground_2024,
            engelbrecht_viability_2018,
            grandi_design_2021,
            grandi_simulating_2020,
            arregui_augmented_2022,
            haque_augmented_2020,
            ii_mixed_2020,
            hu2021human,
            hu_seeing_2022,
            ismael_radiological_2023,
            karakostas_real-time_2024,
            kelley_guiding_2024,
            kim_mobile_2016,
            nalamothu2024leveraging,
            newaz_supporting_2015,
            nilsson_using_2009,
            schmorrow_mixed-methods_2024,
            nunes2019augmented,
            nunes_augmented_2018,
            peretti_augmented_2022,
            phillips2020robotic,
            balfour_next_2012,
            rainer2009simrad,
            safranoglou2024augmented,
            sainidis_single-handed_2021,
            sebillo_training_2016,
            siu_sidebars_2013,
            stefanidi_real-time_2022,
            tadokoro_robocup-rescue_2000,
            tiemann_celidon_2020,
            umlauft_communication_2016,
            walker_mixed_2021,
            wang_person--interest_2013,
            wilchek2025ajna,
            zhang_exploring_2024}
      \\[4pt]

      & Others & 4 &
      \cite{mannuru_mobile_2022,
            mantoro_pathfinding_2021,
            sampson_ar_2022,
            sharma_situational_2020}
      \\

    \bottomrule
  \end{tabularx}
  \normalsize
\end{table*}

%% file: Sections/Tables/requirements_table.tex
\begin{table*}[t]
  \small
  \setlength{\tabcolsep}{6pt}
  \renewcommand{\arraystretch}{1.10}

  \caption{Publications classified by the primary data source that was processed or displayed by the AR System. This facet, termed ``Requirements,'' categorizes systems based on the fundamental information they are designed to handle in support of first responder tasks.}
  \label{table:requirements}
  \centering
  \begin{tabularx}{\textwidth}{@{}l l c X@{}}
    \toprule
    \textbf{Facet} & \textbf{Category} & \textbf{Count} & \textbf{Publications} \\
    \midrule

     \makecell[l]{Requirements} & Physiological & 6 &
      \cite{antevski_5g-based_2021,
            azpiroz2024white,
            garcia_health-5g_2023,
            garcia2024smart,
            umlauft_communication_2016,
            wang_method_2023}
      \\[4pt]

      & Patient/Suspect & 10 &
      \cite{christian_google_2014,
            engelbrecht_viability_2018,
            friedman2024prehospital,
            grandi_simulating_2020,
            arregui_augmented_2022,
            haque_augmented_2020,
            ierache_augmented_2016,
            schlosser_head-worn_displays_2021,
            wang_person--interest_2013,
            wang_method_2023}
      \\[4pt]

      & Environment & 54 &
      \cite{antevski_5g-based_2021,
            azpiroz2024white,
            berndt_optical_2015,
            bhattarai_embedded_2020,
            bram-larbi_collision_2020,
            cai_mobile_2023,
            campos_mobile_2019,
            chae_design_2023,
            chalimas_cross-device_2023,
            chan_erwear_2016,
            cheng_effective_2023,
            chroust2009training,
            crowley_ar_2014,
            dave_augmenting_2013,
            davis2002augmented,
            demirkan_underground_2024,
            dianxi_design_2021,
            engelbrecht_viability_2018,
            del2014red,
            garcia2024smart,
            gkika_object_2023,
            grandi_design_2021,
            grandi_simulating_2020,
            arregui_augmented_2022,
            haque_augmented_2020,
            ii_mixed_2020,
            hu_seeing_2022,
            karakostas_real-time_2024,
            kim_mobile_2016,
            koutitas2019virtual,
            li2019fire,
            majumdar_cloud-based_2023,
            mannuru_mobile_2022,
            mantoro_pathfinding_2021,
            nalamothu2024leveraging,
            newaz_supporting_2015,
            schmorrow_mixed-methods_2024,
            nunes2019augmented,
            oregui2024augmented,
            phillips2020robotic,
            rainer2009simrad,
            safranoglou2024augmented,
            sampson_ar_2022,
            schonauer_3d_2013,
            sebillo_training_2016,
            sharma_situational_2020,
            stefanidi_real-time_2022,
            tadokoro_robocup-rescue_2000,
            tiemann_celidon_2020,
            umlauft_communication_2016,
            weichelt_augmented_2018,
            whitlock_designing_2019,
            wilchek2025ajna,
            zhang_exploring_2024}
      \\[4pt]

      & Context Unaware & 29 &
      \cite{arif_comparative_2019,
            schmorrow_implementing_2016,
            balfour_what_2013,
            brandao_using_2017,
            broach_usability_2018,
            christaki_augmented_2022,
            doswell2020juxtopia,
            follmann2019technical,
            hu2021human,
            ismael_radiological_2023,
            kapalo_comparing_2022,
            kelley_guiding_2024,
            lackey_new_2016,
            lugtenberg_magicbook_2023,
            nelson_user-centered_2022,
            nilsson_using_2009,
            nunes_augmented_2018,
            oconnor_augmented_2023,
            oconnor_pilot_2024,
            pavlopoulos_augmented_1997,
            peretti_augmented_2022,
            balfour_next_2012,
            frasson_cpr_2023,
            rojas-munoz_evaluation_2020,
            sainidis_single-handed_2021,
            schlosser2024effects,
            siu_sidebars_2013,
            stone_mixed_2017,
            walker_mixed_2021}
      \\
    \midrule
    \multicolumn{4}{@{}p{\dimexpr\textwidth-2\tabcolsep}@{}}{\footnotesize Note: ``Context Unaware'' refers to systems where the primary AR interface elements described did not explicitly depend on real-time physiological, specific patient/suspect, or detailed environmental data feeds for their core functionality, or where such data dependencies were not reported as central to the AR features discussed.} \\
    \bottomrule
  \end{tabularx}
  \normalsize
\end{table*}

%% file: Sections/Tables/display_hardware_table.tex
\begin{table*}[t]
  \small
  \setlength{\tabcolsep}{6pt}
  \renewcommand{\arraystretch}{1.10}

  \caption{Publications classified by display hardware.}
  \label{table:display_hardware}
  \centering
  \begin{tabularx}{\textwidth}{@{}l l c X@{}}
    \toprule
    \textbf{Facet} & \textbf{Category} & \textbf{Count} & \textbf{Publications} \\
    \midrule

      \makecell[l]{Display\\ Hardware} & OST HMD & 52 &
      \cite{antevski_5g-based_2021,
            arif_comparative_2019,
            azpiroz2024white,
            berndt_optical_2015,
            bhattarai_embedded_2020,
            brandao_using_2017,
            broach_usability_2018,
            cai_mobile_2023,
            chae_design_2023,
            chalimas_cross-device_2023,
            chan_erwear_2016,
            cheng_effective_2023,
            christaki_augmented_2022,
            christian_google_2014,
            demirkan_underground_2024,
            doswell2020juxtopia,
            del2014red,
            follmann2019technical,
            friedman2024prehospital,
            garcia_health-5g_2023,
            garcia2024smart,
            gkika_object_2023,
            arregui_augmented_2022,
            haque_augmented_2020,
            ii_mixed_2020,
            hu_seeing_2022,
            ismael_radiological_2023,
            kelley_guiding_2024,
            koutitas2019virtual,
            lackey_new_2016,
            li2019fire,
            nalamothu2024leveraging,
            nelson_user-centered_2022,
            newaz_supporting_2015,
            oconnor_augmented_2023,
            oconnor_pilot_2024,
            oregui2024augmented,
            peretti_augmented_2022,
            phillips2020robotic,
            frasson_cpr_2023,
            rojas-munoz_evaluation_2020,
            safranoglou2024augmented,
            sainidis_single-handed_2021,
            schlosser2024effects,
            sharma_situational_2020,
            stone_mixed_2017,
            umlauft_communication_2016,
            wang_person--interest_2013,
            wang_method_2023,
            whitlock_designing_2019,
            wilchek2025ajna,
            zhang_exploring_2024}
      \\[4pt]

      & VST HMD & 10 &
      \cite{schmorrow_implementing_2016,
            cai_mobile_2023,
            grandi_design_2021,
            grandi_simulating_2020,
            hu2021human,
            hu_seeing_2022,
            nilsson_using_2009,
            schonauer_3d_2013,
            stone_mixed_2017,
            walker_mixed_2021}
      \\[4pt]

      & HMD (Unspecified) & 17 &
      \cite{balfour_what_2013,
            chroust2009training,
            davis2002augmented,
            dianxi_design_2021,
            karakostas_real-time_2024,
            lugtenberg_magicbook_2023,
            nunes_augmented_2018,
            mantoro_pathfinding_2021,
            nilsson_using_2009,
            schmorrow_mixed-methods_2024,
            pavlopoulos_augmented_1997,
            rainer2009simrad,
            schlosser_head-worn_displays_2021,
            schonauer_3d_2013,
            stefanidi_real-time_2022,
            tadokoro_robocup-rescue_2000,
            tiemann_celidon_2020}
      \\[4pt]

      & Mobile Device & 22 &
      \cite{cai_mobile_2023,
            campos_mobile_2019,
            crowley_ar_2014,
            dave_augmenting_2013,
            engelbrecht_viability_2018,
            haque_augmented_2020,
            hu_seeing_2022,
            ierache_augmented_2016,
            ismael_radiological_2023,
            kim_mobile_2016,
            majumdar_cloud-based_2023,
            mannuru_mobile_2022,
            mantoro_pathfinding_2021,
            nunes2019augmented,
            nunes_augmented_2018,
            sampson_ar_2022,
            sebillo_training_2016,
            sharma_situational_2020,
            siu_sidebars_2013,
            umlauft_communication_2016,
            weichelt_augmented_2018,
            whitlock_designing_2019}
      \\[4pt]

      & Stationary & 7 &
      \cite{balfour_what_2013,
            bram-larbi_collision_2020,
            dave_augmenting_2013,
            davis2002augmented,
            kapalo_comparing_2022,
            balfour_next_2012,
            umlauft_communication_2016}
      \\

    \bottomrule
  \end{tabularx}
  \normalsize
\end{table*}

%% file: Sections/Tables/display_context_table.tex
\begin{table*}[t]
  \small
  \setlength{\tabcolsep}{6pt}
  \renewcommand{\arraystretch}{1.10}

  \caption{Publications classified by display context.}
  \label{table:display_context}
  \centering
  \begin{tabularx}{\textwidth}{@{}l l c X@{}}
    \toprule
    \textbf{Facet} & \textbf{Category} & \textbf{Count} & \textbf{Publications} \\
    \midrule

      \makecell[l]{Display\\ Context} & Spatial & 76 &
      \cite{antevski_5g-based_2021,
            schmorrow_implementing_2016,
            balfour_what_2013,
            bhattarai_embedded_2020,
            bram-larbi_collision_2020,
            brandao_using_2017,
            cai_mobile_2023,
            campos_mobile_2019,
            chae_design_2023,
            chalimas_cross-device_2023,
            cheng_effective_2023,
            christaki_augmented_2022,
            chroust2009training,
            crowley_ar_2014,
            dave_augmenting_2013,
            davis2002augmented,
            demirkan_underground_2024,
            dianxi_design_2021,
            doswell2020juxtopia,
            engelbrecht_viability_2018,
            del2014red,
            friedman2024prehospital,
            garcia_health-5g_2023,
            garcia2024smart,
            gkika_object_2023,
            grandi_design_2021,
            grandi_simulating_2020,
            arregui_augmented_2022,
            haque_augmented_2020,
            ii_mixed_2020,
            hu2021human,
            hu_seeing_2022,
            ierache_augmented_2016,
            kapalo_comparing_2022,
            karakostas_real-time_2024,
            kelley_guiding_2024,
            kim_mobile_2016,
            koutitas2019virtual,
            lackey_new_2016,
            lugtenberg_magicbook_2023,
            majumdar_cloud-based_2023,
            mannuru_mobile_2022,
            mantoro_pathfinding_2021,
            nalamothu2024leveraging,
            nelson_user-centered_2022,
            nilsson_using_2009,
            schmorrow_mixed-methods_2024,
            nunes2019augmented,
            nunes_augmented_2018,
            oregui2024augmented,
            pavlopoulos_augmented_1997,
            peretti_augmented_2022,
            phillips2020robotic,
            balfour_next_2012,
            rainer2009simrad,
            frasson_cpr_2023,
            rojas-munoz_evaluation_2020,
            safranoglou2024augmented,
            sainidis_single-handed_2021,
            sampson_ar_2022,
            schlosser_head-worn_displays_2021,
            schlosser2024effects,
            schonauer_3d_2013,
            sebillo_training_2016,
            sharma_situational_2020,
            siu_sidebars_2013,
            stefanidi_real-time_2022,
            stone_mixed_2017,
            tiemann_celidon_2020,
            walker_mixed_2021,
            wang_person--interest_2013,
            wang_method_2023,
            weichelt_augmented_2018,
            whitlock_designing_2019,
            wilchek2025ajna,
            zhang_exploring_2024}
      \\[4pt]

      & HUD & 40 &
      \cite{arif_comparative_2019,
            azpiroz2024white,
            schmorrow_implementing_2016,
            balfour_what_2013,
            berndt_optical_2015,
            brandao_using_2017,
            broach_usability_2018,
            campos_mobile_2019,
            chae_design_2023,
            chalimas_cross-device_2023,
            chan_erwear_2016,
            cheng_effective_2023,
            christian_google_2014,
            crowley_ar_2014,
            doswell2020juxtopia,
            follmann2019technical,
            garcia_health-5g_2023,
            garcia2024smart,
            grandi_design_2021,
            arregui_augmented_2022,
            haque_augmented_2020,
            ierache_augmented_2016,
            kim_mobile_2016,
            li2019fire,
            newaz_supporting_2015,
            nunes2019augmented,
            nunes_augmented_2018,
            oconnor_augmented_2023,
            oconnor_pilot_2024,
            oregui2024augmented,
            sampson_ar_2022,
            schlosser_head-worn_displays_2021,
            sebillo_training_2016,
            siu_sidebars_2013,
            tiemann_celidon_2020,
            umlauft_communication_2016,
            wang_person--interest_2013,
            wang_method_2023,
            weichelt_augmented_2018,
            zhang_exploring_2024}
      \\[4pt]

      & Unreported & 2 &
      \cite{ismael_radiological_2023,
            tadokoro_robocup-rescue_2000}
      \\

    \bottomrule
  \end{tabularx}
  \normalsize
\end{table*}

%% file: Sections/3.3analysis.tex
\subsection{Analysis of AR Interface Elements}\label{sec:analysis}

This section presents the analysis of the AR Interface facet of the reviewed literature, focusing on the specific techniques and methods identified. In this work, we use the term \textbf{Interface Element} to denote the granular components, or building blocks, of AR User Interfaces designed for first responders. The following sub-sections detail each Interface Element identified. The whole data  for papers mentioning them can be found in Table \ref{table:longtable}.

\subsubsection{Environment Awareness}
Interfaces in this category aim to enhance the first responder's understanding of their physical surroundings, effectively supporting the perception, comprehension, or projection levels of Situational Awareness (SA)~\cite{endsley1995toward}.

\vspace{3pt}\noindent\textbf{Edge Detection:} Edge detection interfaces are spatial augmentations designed to enhance perception in low visibility conditions, such as smoke or fog, by superimposing outlines of physical objects. This directly aids SA perception by making environmental elements more discernible. Two papers addressed Edge Detection, Dianxi et al.~\cite{dianxi_design_2021} proposed its use for Firefighters in Search and Rescue, stating that ``the AR augmented reality of the terminal processor presents an imaging higher than reality,'' suggesting a perceptual benefit derived from this simplification. More concretely, Demirkan et al.~\cite{demirkan_underground_2024} evaluated an Edge Detection system using the Hololens for Search and Rescue in low visibility. Their findings demonstrated effectiveness, reporting that Edge Detection significantly improved search and rescue time and enhanced the ability to detect unexpected hazards, likely due to the clearer delineation of the environment.

\vspace{3pt}\noindent\textbf{X-Ray:}
X-Ray vision, as an AR interface element, provides sight through occluding surfaces by overlaying information about obscured objects. This directly enhances perception by revealing otherwise hidden elements and can contribute to comprehension by clarifying spatial relationships or the status of occluded items. The primary design goal is to extend the user's perceptual field, supporting tasks such as target acquisition, hazard identification, or procedural guidance. Five papers discussed this concept, though the literature offers limited direct evidence from comparative user studies on its effectiveness (RQ2).

Applications varied by discipline: for firefighters, X-Ray was evaluated to view simulated victims in obscured environments~\cite{chae_design_2023} or proposed to visualize teammates and navigation lines through obstructions to aid coordination and route planning~\cite{grandi_design_2021}. For EMS, it was implemented to overlay patient anatomy for airway management procedures, aiming for improved procedural accuracy through in-situ visualization~\cite{davis2002augmented}. In law enforcement, X-Ray was implemented to provide perspective through walls for tactical awareness~\cite{phillips2020robotic}. It was also discussed as a method to maintain awareness of occluded Points of Interest (PoIs)~\cite{brandao_using_2017}. While X-Ray vision is a recurring concept with clear motivations for enhancing perception, more research is needed to demonstrate its practical effectiveness and address design challenges (e.g., visual clutter, depth perception) in operational first responder tasks.

\input{Sections/Tables/output_channel_table}

\vspace{3pt}\noindent\textbf{Spatial Reconstruction:}
Spatial Reconstruction involves displaying a virtual model (e.g., outline, wire-frame, hologram) of a structure or object, often to provide insight into occluded areas or to aid in understanding complex environments and layouts, particularly when direct visibility is compromised or the environment is too hazardous to explore physically. This supports SA comprehension by offering a holistic view of complex space and can aid SA projection improving planning of entry or movement. Four papers in this review featured Spatial Reconstruction.

The literature primarily describes interface designs or system implementations rather than comparative evaluations of user performance. For example, Howard et al.~\cite{ii_mixed_2020} proposed a system for disaster scenarios where a wire-frame model of the original structure of a destroyed building was spatially augmented to assist first responders understand the rubble, while Kim et al.~\cite{kim_mobile_2016} described an implementation using sensors to create 3D models of buildings post-disaster for damage assessment. While the utility of such systems is evident, the papers did not present specific studies demonstrating or quantifying the effectiveness of these spatial reconstruction interfaces on first responder task performance or situational awareness compared to traditional methods or other AR approaches.

\vspace{3pt}\noindent\textbf{Gaze Indicators:} 
Gaze indicators are visual cues designed to direct a user's attention to a specific location or object, often leveraging principles of pre-attentive visual processing to rapidly guide the eye. These directly support SA perception by ensuring critical elements are noticed. Kelley et al.~\cite{kelley_guiding_2024} provided the sole direct evaluation of this element in the review. They compared gaze cue types, represented by gaze lines (a line from the user to the object), gaze wedges (a triangle pointing towards the object), and gaze arrows (a 3D spatially augmented arrow). All types of gaze cues improved visual search time compared to no cue, with gaze lines was identified as the most effective, reportedly reducing search time by 5 seconds and improving accuracy by 12\%. This study provides clear evidence for the effectiveness of gaze indicators, particularly gaze lines, in guiding visual search for first responders, likely by reducing the cognitive effort required to locate targets in complex scenes.

\subsubsection{Object Awareness}
Interfaces in this category focus on drawing attention to and providing information about specific objects or entities within the environment by making objects salient and/or by providing contextual information about them.

\vspace{3pt}\noindent\textbf{Object Highlighting:}
Object Highlighting involves superimposing visual cues (e.g., color, bounding boxes, outlines) onto objects of importance within a scene. A core design challenge is to draw attention effectively without inducing clutter or cognitive overload. This was a frequently discussed Interface Element, appearing in 20 papers. Regarding its demonstrated effectiveness (RQ2), the literature offers some insights, although many papers focus primarily on system proposals or implementations.

A notable evaluation by Ntoa et al.~\cite{schmorrow_mixed-methods_2024} examined the DARLENE system, an AR framework for law enforcement using an OST HMD. DARLENE employs a Deep Neural Network for object/person identification, highlighting them with a shaded box, label, and confidence level. Ntoa et al.'s experiment, assessing situational awareness, workload, and user experience under stress and non-stress conditions, found that DARLENE (which heavily features object highlighting) improves situational awareness under stress without increasing perceived workload. This provides evidence for the effectiveness of such highlighting systems in high-stakes scenarios. Within the DARLENE framework, Stefanidi et al.~\cite{stefanidi_real-time_2022} proposed adaptive highlighting based on detail levels and color, addressing the need to manage information density. Karakostas et al.~\cite{karakostas_real-time_2024} described specific color-coding (white, yellow, red, blue) for highlighting people (e.g., victim, suspect, ally), leveraging semantic color associations to convey status rapidly. However, specific evaluations of these color schemes' effectiveness were not detailed.

User feedback also contributes to understanding effectiveness and critical design aspects. Zhang et al.~\cite{zhang_exploring_2024} conducted a structured interview study with first responders evaluating a system with object highlighting (contours and shading). Participants emphasized that effective highlighting cues should help them ``understand the size and shape of important objects and better determine subsequent actions.'' They also suggested that targeted highlighting, such as augmenting a doorknob and opening direction rather than the entire door, ``could be more important,'' indicating that the granularity and specificity of the highlighting technique are crucial design considerations that significantly impact perceived utility and effectiveness.

Other papers described various applications of object highlighting where effectiveness was implied rather than empirically demonstrated. These include highlighting potential evidence or suspects in crime scene investigations~\cite{haque_augmented_2020}, using color-coding to indicate severity levels of scanned items (e.g., driver's licenses, illegal items)~\cite{grandi_simulating_2020}, combining highlighting with X-Ray views to give perspective on items through walls~\cite{phillips2020robotic}, visualizing patient temperature data for EMS via color-coded bounding boxes~\cite{wang_method_2023}, and employing techniques like color-shading for people or outlining windows and doors for firefighters~\cite{bhattarai_embedded_2020}. These examples illustrate diverse applications and design choices (e.g., color, shape, level of detail) in object highlighting, but further research is needed to systematically evaluate the effectiveness of these different techniques across various first responder tasks and contexts, particularly concerning cognitive load and task performance.

\subsubsection{Navigation}
Navigation-specific interfaces are crucial for first responders, aiding them in orienting within and moving through complex, often unfamiliar or hazardous environments. These interfaces primarily support SA comprehension of the spatial layout and SA projection by facilitating route planning and goal-directed movement. Common elements include maps, compasses, radars, points of interest (PoIs), breadcrumbs, and directional arrows/lines. These were frequently mentioned (103 instances across various disciplines), indicating the high importance of navigational AR interfaces for first responders.

\vspace{3pt}\noindent\textbf{``Traditional'' Navigation:} This group includes established navigation tools adapted for AR.
Maps, both 2D and 3D, are foundational, aiming to provide a cognitive model of the environment, crucial  for route planning. 39 papers discussed maps.
\textit{2D Maps} (29 papers) present a top-down view, where the main design challenge is balancing information density and legibility. Heading-up displays were found to be more effective than north-up by reducing mental rotation~\cite{smets_effects_2008}. Implementations included displaying 2D maps on flat surfaces with 3D augmented objects~\cite{peretti_augmented_2022} or as interactive, annotatable displays for shared SA~\cite{nilsson_using_2009}, with users preferring them over paper maps due to reduced clutter.
\textit{3D Maps} (10 papers), often spatially augmented, aim for intuitive representation of three-dimensional spaces to aid comprehension of verticality and complex structures, though they can introduce occlusion or cognitive load. One study found no significant difference in perceived workload between 2D and 3D maps for firefighters~\cite{kapalo_comparing_2022}, suggesting that workload-related benefits are not universal.

Compasses and Radars provide immediate orientation and awareness of nearby entities.
\textit{Compasses} (e.g., ~\cite{chalimas_cross-device_2023, nunes2019augmented, nunes_augmented_2018, peretti_augmented_2022, weichelt_augmented_2018}) offer directional orientation, supporting perception of current heading and comprehension of spatial relationships, often used to orient towards PoIs~\cite{nunes2019augmented, nunes_augmented_2018, weichelt_augmented_2018, campos_mobile_2019}.
\textit{Radars} (e.g.,~\cite{azpiroz2024white, campos_mobile_2019, garcia2024smart, kim_mobile_2016, nunes_augmented_2018, oregui2024augmented}) offer a schematic overview of nearby entities (teammates, hazards, objectives) not in direct view, supporting perception and comprehension of the immediate vicinity.
For both compasses and radars, the literature often described them as components within larger systems, with limited specific evaluations of their standalone effectiveness.

\vspace{3pt}\noindent\textbf{``Neo-'' Navigation:} This category encompasses more dynamic or less conventional AR navigation aids.
\textit{Breadcrumbs} track the user's path to offload path memory. Proposals included color-coding for temporal context~\cite{schmorrow_implementing_2016} or path sharing~\cite{christaki_augmented_2022}. A non-AR study highlighted potential usability issues with dot-based breadcrumbs due to inaccuracy, suggesting lines might be better~\cite{wilson_head-mounted_2009}.

\textit{Points of Interest (PoIs)} mark important locations by highlighting them and associating them with relevant information. Managing PoI visibility and information content to avoid clutter is a key design challenge. PoIs were prevalent (41 papers). Evaluated systems showed user preference for AR maps with PoIs over paper maps because of reduced clutter and improved SA~\cite{nilsson_using_2009}. Techniques to manage information density included clustering PoIs by location with expandable details, using sidebars for off-screen PoIs, and integrating PoIs with compasses for orientation~\cite{campos_mobile_2019, siu_sidebars_2013}. PoIs can be placed through various mechanisms, including remote explorers~\cite{christaki_augmented_2022}, command centers, or automated systems~\cite{dave_augmenting_2013}. While widely adopted in designs and implementations, more empirical evidence on the effectiveness of different PoI designs and interaction techniques would be beneficial.

\textit{Navigation Arrows/Lines} provide direct, unambiguous directional guidance, reducing map interpretation and supporting perception of the immediate path. Examples include arrows on 3D maps for
evacuation~\cite{sharma_situational_2020} or as part of simulated AR systems~\cite{schmorrow_implementing_2016}. However, the literature offers limited direct comparative evaluations of their effectiveness against other methods, and over-reliance remains a potential concern.

\subsubsection{Alerts}
Alerts are interface elements designed to warn first responders or draw their attention to critical information regarding context or objects by ensuring critical cues are noticed, and by conveying the 
\makebox[\linewidth][s]{nature and urgency of the information. A key design challenge for}

\onecolumn
\input{Sections/Tables/longtable}
\begin{multicols}{2}  
\noindent alerts is to be conspicuous and informative enough to be noticed and understood in high-stress situations, yet subtle enough to avoid causing undue distraction or alarm. While alerts can be multimodal (haptic, auditory), this review focuses on those with a visual component. In total, 12 papers featured alert interfaces, with a notable concentration (10 papers) targeting law enforcement, often leveraging sensor data and object highlighting to enhance situational awareness.

The DARLENE system, which incorporates visual alerts with object highlighting and icons, has been shown to improve situational awareness for law enforcement under stress~\cite{stefanidi_real-time_2022, karakostas_real-time_2024, schmorrow_mixed-methods_2024}. User feedback from studies like Zhang et al.~\cite{zhang_exploring_2024} also provides insights into perceived effectiveness, with participants emphasizing the necessity of head vibration for effective alerts in various first responder scenarios. However, many other descriptions of alert systems focus on their proposed functionality rather than presenting rigorous comparative evaluations of their effectiveness.

\vspace{3pt}\noindent\textbf{Object Awareness Alerts:} These alerts aim to enhance awareness of specific objects or entities, often by combining highlighting with explicit warning cues, making objects salient and providing immediate status or threat information. The design rationale frequently involves leveraging multiple sensory channels or distinct visual properties to ensure critical information is perceived. Examples include systems using visual (e.g., colored haze) and haptic alerts prioritized by notification severity for suspicious items~\cite{grandi_simulating_2020, grandi_design_2021}, or facial recognition with bounding box highlighting for wanted suspects~\cite{wang_person--interest_2013}. The DARLENE system's alert widget combined object  
highlighting, symbolic icons (e.g., gun, knife), and semantic color-coding (e.g., orange, red) to indicate danger, reportedly improving SA~\cite{stefanidi_real-time_2022, karakostas_real-time_2024, schmorrow_mixed-methods_2024}. Similar approaches using object highlighting and warning symbols were proposed by Cheng et al.~\cite{cheng_effective_2023}. For EMS, a Google Glass system used QR codes for object recognition to display hazard information (e.g., substance type) and suggested actions with warning symbols~\cite{berndt_optical_2015}. These examples show a focus on conveying critical object-related information, with DARLENE providing the most direct evidence of effectiveness in enhancing SA. 

\vspace{3pt}\noindent\textbf{Environmental Awareness Alerts:}
These alerts focus on environmental hazards like heat, smoke, or physiological stress, with the primary design goal of preventing harm to the first responder or managing resources by providing timely warnings about ambient conditions or the responder's own state. They directly support perception of environmental conditions and comprehension of associated risks, which can inform SA projection regarding potential future states or necessary preventative actions. Four of the 12 alert-focused papers targeted firefighters with such systems. Regarding effectiveness (RQ2), Zhang et al.~\cite{zhang_exploring_2024} found through user evaluations that head vibration was considered necessary for effective environmental alerts, such as for extreme conditions for firefighters or suspicious activity for law enforcement/EMS, suggesting that non-visual channels can be crucial when visual attention is saturated. Grandi et al.~\cite{grandi_design_2021} proposed a color-coded haze in the user's periphery as an alert to environmental danger, a design choice aimed at intuitive, less obtrusive peripheral perception, though its effectiveness was not empirically demonstrated. While the intent of these systems is to improve safety and awareness of environmental conditions, more specific evaluations comparing different alert modalities (e.g., visual intensity, haptic patterns, auditory cues) and their impact on first responder performance and safety are needed.

\subsubsection{On Demand Interfaces}
On-Demand Interfaces are dynamic spatially augmented interfaces that remain hidden until explicitly called upon by the user. The primary design goal is to reduce persistent visual clutter and associated cognitive load, presenting information or controls only when contextually relevant or actively sought by the user. By managing information flow and minimizing distractions, these interfaces can indirectly support overall SA by freeing up cognitive resources for critical tasks. Five papers in this review featured such interfaces. The literature primarily describes their implementation and intended benefits (e.g., minimizing cognitive load by showing information only when needed) rather than providing direct comparative evaluations against always-visible interfaces or other interaction paradigms.

\vspace{3pt}\noindent\textbf{Body-Attached:}
This approach involves interfaces metaphorically attached to the user's body, like a virtual wristwatch. The design rationale often leverages proprioception and familiar physical interaction paradigms (e.g., looking at one's wrist) to make access intuitive. Their impact on SA is primarily through efficient access to information that supports SA processes, rather than directly providing SA cues themselves. Grandi et al.~\cite{grandi_design_2021, grandi_simulating_2020} described an arm-mounted ``wrist-watch'' interface for information relay, using a physical device and haptics for alerts. The implied benefit is an intuitive and quickly accessible information display, though its effectiveness compared to other on-demand or HUD-based displays was not empirically evaluated.

\vspace{3pt}\noindent\textbf{Summoned:}
Summoned interfaces, typically hand menus, appear when triggered by a specific gesture. The design goal is to provide quick, potentially eyes-free access to controls or secondary information without continuous visual obstruction, thereby allowing users to maintain perception of their environment while accessing tools. García et al.~\cite{garcia_health-5g_2023} and Nelson et al.~\cite{nelson_user-centered_2022} described hand menus activated by turning the palm up, used for tasks like UI control~\cite{garcia_health-5g_2023} or a specialized triage tag with a body chart for annotation~\cite{nelson_user-centered_2022}. Christaki et al.~\cite{christaki_augmented_2022} also used a hand menu for placing PoIs. However, challenges with summoned interfaces can include the reliability and discoverability of activation gestures, especially under stress or with gloved hands. The effectiveness of these
\makebox[\linewidth][s]{gestural summoning methods (e.g., reliability, speed, cognitive}

\end{multicols}
\twocolumn
\noindent load, preference) compared to other interaction methods for accessing secondary information was not a primary focus of the evaluations covered in this review. Further research could explore the usability and performance benefits of different on-demand interface types and summoning techniques in first responder contexts.

\subsubsection{Augmented Guidance}                                                    
Augmented Guidance encompasses AR Interface Elements and systems primarily designed to direct, instruct, or enhance a first responder's understanding and execution of tasks. The core design principle is to reduce cognitive load associated with complex procedures or information recall, and to improve task accuracy and efficiency by providing contextual, step-by-step, or enriched information directly within the user's field of view or interaction space. These interfaces actively guide the user, rather than merely presenting data on tasks and procedures. This review categorizes Augmented Guidance into two main types: \textbf{Augmented Support}, which provides direct assistance for procedures and information comprehension, and \textbf{Augmented Enhancement}, which enriches the user's perception or capabilities regarding specific aspects of the environment or subjects of interest. In the reviewed literature, 27 papers used some form of augmented guidance.

\vspace{3pt}\noindent\textbf{Augmented Support:}
This sub-category includes interfaces that directly assist first responders in performing procedures or comprehending information through step-by-step instructions or explanatory visual aids, directly contributing to the comprehension of the task at hand.
                                                                                      
\textit{Checklists} are interfaces designed to guide users step-by-step through procedures, ensuring all procedural steps are addressed and action sequence is properly structured. The primary design rationale for AR checklists is to ensure procedural adherence, reduce errors of omission, and standardize the execution of complex or infrequently performed tasks, especially in high-stress environments where memory recall can be fallible. Examples include voice-interactive checklist interfaces on Google Glass for guiding triage~\cite{follmann2019technical, berndt_optical_2015}, and HUD-based checklists on OST HMDs for emergency medical procedures like cricothyroidotomy, which aimed to assess proficiency with just-in-time AR guidance~\cite{oconnor_augmented_2023, oconnor_pilot_2024}. Checklists have also been combined with graphics and remote expert guidance, with evaluations touching on task performance and communication~\cite{schlosser2024effects}. These studies suggest AR checklists are perceived as beneficial for procedural guidance, particularly for complex or infrequently performed tasks. However, more extensive comparative studies are needed to robustly quantify their effectiveness.                                                                                      
\textit{Annotations}, broadly defined as explanatory notes or visual markings, are versatile Interface Elements. In the context of Augmented Support, annotations are specifically those visual cues--such as tooltips, drawings, icons, or explanatory figures--that are overlaid to clarify information, highlight important details, or guide users through procedures. A key design challenge is to ensure they are informative and relevant without creating visual clutter. Sixteen papers featured annotations in a supportive guidance role.

\textit{Tool tip annotations} provide contextual information about actions, interfaces, or objects, by delivering just-in-time cues precisely where needed. Hu et al.~\cite{hu2021human} demonstrated their effectiveness in a simulated disaster victim extraction, where tooltips significantly reduced search time and false alarms, and increased victim discovery.

\textit{Drawn/written annotations}, often used in telementoring, allow specialists to draw or write directly into the first responder's field of view, conveying expert knowledge and indicating correct actions. This leverages human expertise for dynamic situations where pre-programmed guidance may be insufficient. Such systems were proposed or implemented for tasks like EMS victim extraction~\cite{pavlopoulos_augmented_1997}, or for medical procedures involving AR markers or drawings projected onto a patient's body~\cite{schlosser_head-worn_displays_2021,lackey_new_2016}. Rojas-Muñoz et al.~\cite{rojas-munoz_evaluation_2020} found AR telementoring with drawn annotations for cricothyroidotomies non-inferior to in-person guidance and superior to audio-only, demonstrating effectiveness for complex procedural guidance.

\textit{Graphical annotations} (gauges, charts, icons) summarize information visually, translating complex data into easily digestible formats. Examples include HUD-based gauges for firefighters displaying environmental data (temperature, oxygen, CO levels)~\cite{cheng_effective_2023}, and graphical body charts with color-coded triage status and treatment timelines for EMS  
\cite{arregui_augmented_2022, nelson_user-centered_2022}. Graphical images have also been combined with checklists in remote mentoring systems~\cite{schlosser2024effects}. While the effectiveness of specific graphical annotation types often relies on intuitive design, more targeted studies comparing different styles would be beneficial.

Overall, annotations, particularly tooltips and drawn/written annotations in telementoring, have shown demonstrated effectiveness in improving task performance and procedural accuracy. The effectiveness of graphical annotations is often implied by their design for enhanced information assimilation, but more direct comparative studies would strengthen these claims.                                                
                                                                                      
\vspace{3pt}\noindent\textbf{Augmented Enhancement:}                                  
This sub-category includes interfaces that enrich a first responder's perception or capabilities concerning specific elements of their environment or subjects of interest. The core design principle is to provide access to information or views not otherwise perceptible, effectively extending the user's sensory or cognitive reach to improve understanding, safety, or operational capacity.

\textit{Vitals Monitors} provide a virtual recreation of traditional medical monitoring displays, aiming to enhance a first responder's awareness of a patient's condition. The design rationale is to offer hands-free, continuous access to critical patient data. Implementations included multiple display versions on HoloLens (movable window, on-demand menu, HUD) presenting raw data and trends~\cite{garcia_health-5g_2023}, HMD-based proposals~\cite{schlosser_head-worn_displays_2021}, and spatially augmented summarized data~\cite{antevski_5g-based_2021}. While these approaches aim to improve SA by making patient data more accessible, the summarized works did not present specific comparative evaluations of their effectiveness against traditional methods. Thus, the practical effectiveness of AR vitals monitors, while intuitively appealing, warrants more rigorous empirical investigation.

\textit{Live Video Feeds}, often integrated as part of telepresence or remote sensing systems, provide users with real-time visual information from other locations (e.g., drone cameras, fixed surveillance cameras). The design goal is to enhance remote SA, provide views into hazardous areas, and support collaborative decision-making. Potential challenges include managing bandwidth, latency, cognitive load from multiple feeds, and integrating 2D video into 3D perception. The DARLENE system included a Live Feed widget for law enforcement, contributing to overall SA improvements under stress~\cite{karakostas_real-time_2024, schmorrow_mixed-methods_2024}. Other examples include integrating live camera feeds onto virtual maps~\cite{balfour_next_2012} or displaying drone camera feeds in spatially augmented windows~\cite{christaki_augmented_2022, sainidis_single-handed_2021}. While these systems aim to enhance remote perception, more targeted research is needed to quantify their specific contributions to first responder effectiveness.

Other forms of Augmented Enhancement include environmental simulation tools and self-monitoring interfaces. Environmental simulation tools, such as for real-time water level changes in flood scenarios~\cite{safranoglou2024augmented}, aim to support SA projection by enabling predictive analysis and decision-making. Self-monitoring interfaces displaying responder vitals and environmental data~\cite{oregui2024augmented, azpiroz2024white} support SA perception and comprehension of the responder's own state and immediate environmental impact, aiming to enhance safety and endurance. While these systems show promise, further studies quantifying improvements in decision quality, safety outcomes, or operational endurance would be beneficial.

%% file: Sections/Tables/output_channel_table.tex
\begin{table*}[t]
  \small
  \setlength{\tabcolsep}{6pt}
  \renewcommand{\arraystretch}{1.10}

  \caption{Publications classified by output modality.}
  \label{table:output}
  \centering
  \begin{tabularx}{\textwidth}{@{}l l c X@{}}
    \toprule
    \textbf{Facet} & \textbf{Category} & \textbf{Count} & \textbf{Publications} \\
    \midrule

      \makecell[l]{Output\\ Modality} & Visual & 90 &
      \cite{antevski_5g-based_2021,
            arif_comparative_2019,
            azpiroz2024white,
            schmorrow_implementing_2016,
            balfour_what_2013,
            berndt_optical_2015,
            bhattarai_embedded_2020,
            bram-larbi_collision_2020,
            brandao_using_2017,
            broach_usability_2018,
            cai_mobile_2023,
            campos_mobile_2019,
            chae_design_2023,
            chalimas_cross-device_2023,
            chan_erwear_2016,
            cheng_effective_2023,
            christaki_augmented_2022,
            christian_google_2014,
            chroust2009training,
            crowley_ar_2014,
            dave_augmenting_2013,
            davis2002augmented,
            demirkan_underground_2024,
            dianxi_design_2021,
            doswell2020juxtopia,
            engelbrecht_viability_2018,
            del2014red,
            follmann2019technical,
            friedman2024prehospital,
            garcia_health-5g_2023,
            garcia2024smart,
            gkika_object_2023,
            grandi_design_2021,
            grandi_simulating_2020,
            arregui_augmented_2022,
            haque_augmented_2020,
            ii_mixed_2020,
            hu2021human,
            hu_seeing_2022,
            ierache_augmented_2016,
            ismael_radiological_2023,
            kapalo_comparing_2022,
            karakostas_real-time_2024,
            kelley_guiding_2024,
            kim_mobile_2016,
            koutitas2019virtual,
            lackey_new_2016,
            li2019fire,
            lugtenberg_magicbook_2023,
            majumdar_cloud-based_2023,
            mannuru_mobile_2022,
            mantoro_pathfinding_2021,
            nalamothu2024leveraging,
            nelson_user-centered_2022,
            newaz_supporting_2015,
            nilsson_using_2009,
            schmorrow_mixed-methods_2024,
            nunes2019augmented,
            nunes_augmented_2018,
            oconnor_augmented_2023,
            oconnor_pilot_2024,
            oregui2024augmented,
            pavlopoulos_augmented_1997,
            peretti_augmented_2022,
            phillips2020robotic,
            balfour_next_2012,
            rainer2009simrad,
            frasson_cpr_2023,
            rojas-munoz_evaluation_2020,
            safranoglou2024augmented,
            sainidis_single-handed_2021,
            sampson_ar_2022,
            schlosser_head-worn_displays_2021,
            schlosser2024effects,
            schonauer_3d_2013,
            sebillo_training_2016,
            sharma_situational_2020,
            siu_sidebars_2013,
            stefanidi_real-time_2022,
            stone_mixed_2017,
            tadokoro_robocup-rescue_2000,
            tiemann_celidon_2020,
            umlauft_communication_2016,
            walker_mixed_2021,
            wang_person--interest_2013,
            wang_method_2023,
            weichelt_augmented_2018,
            whitlock_designing_2019,
            wilchek2025ajna,
            zhang_exploring_2024}
      \\[4pt]

      & Auditory & 10 &
      \cite{schmorrow_implementing_2016,
            follmann2019technical,
            garcia_health-5g_2023,
            haque_augmented_2020,
            newaz_supporting_2015,
            rojas-munoz_evaluation_2020,
            schlosser2024effects,
            stone_mixed_2017,
            wang_method_2023,
            zhang_exploring_2024}
      \\[4pt]

      & Haptic & 2 &
      \cite{cheng_effective_2023,
            grandi_simulating_2020}
      \\

    \bottomrule
  \end{tabularx}
  \normalsize
\end{table*}

%% file: Sections/Tables/longtable.tex
\begingroup
\scriptsize
\begin{longtable}{@{}p{2.5em}  %
                  p{0.35\textwidth}  %
                  p{0.1\textwidth}  %
                  p{0.05\textwidth} %
                  p{0.37\textwidth}@{}} %

\caption{Summary of Publications and the Interface Elements They Introduce or Analyze}
\label{table:longtable}\\
\toprule

\textbf{Ref} & \textbf{Title} & \textbf{Authors} & \textbf{Year} & \textbf{Interface elements}\\
\midrule
\endfirsthead

\multicolumn{5}{@{}l}{\small\itshape Table~\ref{table:longtable} – continued from previous page}\\
\toprule
\textbf{Ref} & \textbf{Title} & \textbf{Authors} & \textbf{Year} & \textbf{Interface elements}\\
\midrule
\endhead

\midrule
\multicolumn{5}{r@{}}{\small\itshape Continued on next page}\\
\endfoot

\bottomrule
\endlastfoot
    \cite{antevski_5g-based_2021} & A 5G-Based eHealth Monitoring and Emergency Response System: Experience and Lessons Learned & Antevski et al. & 2021  & Points of Interest, Vitals Monitor \\
    \cite{arif_comparative_2019} & A Comparative Study of Rendering Devices for Safety-Critical Applications in Operative Control Rooms & Arif et al. & 2019  & 2D Map, Annotations \\
    \cite{azpiroz2024white} & White Paper on Adaptive Situational Awareness Enhancing Augmented Reality Interface Design on First Responders in Rescue Tasks & Azpiroz et al. & 2024  & Radar, Augmented Enhancement \\
    \cite{schmorrow_implementing_2016} & Implementing user-centered methods and virtual reality to rapidly prototype augmented reality tools for firefighters & Bailie et al. & 2016  & Breadcrumbs, Navigation Arrow \\
    \cite{balfour_what_2013} & The what, why and how of achieving urban telepresence & Balfour and Donnelly & 2013  & Augmented Enhancement \\
    \cite{berndt_optical_2015} & Optical head-mounted displays in mass casualty incidents: Keeping an eye onpatients and hazardous materials & Berndt et al. & 2015  & Alert, Augmented Support \\
    \cite{bhattarai_embedded_2020} & An embedded deep learning system for augmented reality in firefighting applications & Bhattarai et al. & 2020  & Object Highlighting \\
    \cite{bram-larbi_collision_2020} & Collision Avoidance Head-Up Display: Design Considerations for Emergency Services’ Vehicles & Bram-Larbi et al. & 2020  & Navigation Arrow \\
    \cite{brandao_using_2017} & Using augmented reality to improve dismounted operators' situation awareness & Brandao and Pinho & 2017  & Navigation Line, Points of Interest, 2D Map, X-Ray, Spatial Reconstruction \\
    \cite{broach_usability_2018} & Usability and reliability of smart glasses for secondary triage during mass casualty incidents & Broach et al. & 2018  & Annotations \\
    \cite{cai_mobile_2023} & Mobile Incident Command Dashboard (MIS-D) & Cai and Siegel & 2023  & 3D Map, 2D Map, Points of Interest \\
    \cite{campos_mobile_2019} & Mobile Augmented Reality Techniques for Emergency Response & Campos et al. & 2019  & Points of Interest, 2D Map, Radar \\
    \cite{chae_design_2023} & Design of a Mixed Reality System for Simulating Indoor Disaster Rescue & Chae et al. & 2023  & 2D Map, Navigation Arrow, X-Ray, Points of Interest \\
    \cite{chalimas_cross-device_2023} & Cross-Device Augmented Reality Systems for Fire and Rescue based on Thermal Imaging and Live Tracking & Chalimas and Mania & 2023  & 2D Map, Navigation Arrow, Compass \\
    \cite{chan_erwear_2016} & ERWear: Wearable System Design through the Lens of First Responders & Chan et al. & 2016  & 2D Map, Points of Interest \\
    \cite{cheng_effective_2023} & Effective Navigation for an Emergency Response for Firefighters with Wearable Haptic Displays Design & Cheng et al. & 2023  & 2D Map, Points of Interest, Annotations, Object Highlighting, Alert \\
    \cite{christaki_augmented_2022} & Augmented reality points of interest for improved first responder situational awareness & Christaki et al. & 2022  & 2D Map, Points of Interest, On Demand Menu, Breadcrumbs, Live Video Feed, Navigation Arrow, Augmented Enhancement \\
    \cite{christian_google_2014} & Google glass for public safety: Leveraging Google Glass for automatic information retrieval and notification by public safety officers in the field & Christian et al. & 2014  & Alert \\
    \cite{chroust2009training} & Training and supporting first responders by mixed reality environments & Chroust et al. & 2009  & Points of Interest \\
    \cite{crowley_ar_2014} & AR Browser for Points of Interest in Disaster Response in UAV Imagery & Crowley et al. & 2014  & 2D Map, Points of Interest \\
    \cite{dave_augmenting_2013} & Augmenting situational awareness for first responders using social media as a sensor & Dave et al. & 2013  & Points of Interest \\
    \cite{davis2002augmented} & Augmented reality and training for airway management procedures & Davis et al. & 2002  & X-Ray \\
    \cite{demirkan_underground_2024} & Underground mine emergency evacuation planning: AR implementation and casestudy & Demirkan et al. & 2024  & Edge Detection \\
    \cite{dianxi_design_2021} & Design of Intelligent Firefighter Helmet Based on AR Technology & Dianxi et al. & 2021  & Edge Detection \\
    \cite{doswell2020juxtopia} & Juxtopia®CAMMRAD Prepare: A Wearable AI-AR Platform for Clinical Training Emergency First Response Teams & Doswell et al. & 2020  & Augmented Support \\
    \cite{engelbrecht_viability_2018} & Viability of Augmented Content for Field Policing & Engelbrecht and Lukosch & 2018  & Points of Interest, Alert \\
    \cite{del2014red} & Red Cross Triage App Design for Augmented Reality Glasses & Fernández et al. & 2014  & 2D Map, Augmented Support \\
    \cite{follmann2019technical} & Technical support by smart glasses during a mass casualty incident: A randomized controlled simulation trial on technically assisted triage and telemedical app use in disaster medicine & Follmann et al. & 2019  & Augmented Support \\
    \cite{friedman2024prehospital} & Prehospital Pediatric Emergency Training Using Augmented Reality Simulation: A Prospective, Mixed Methods Study & Friedman et al. & 2024  & Augmented Enhancement \\
    \cite{garcia_health-5g_2023} & Health-5G: A Mixed Reality-Based System for Remote Medical Assistance in Emergency Situations & García et al. & 2023  & Vitals Monitor, On Demand Menu, Augmented Enhancement, Augmented Support \\
    \cite{garcia2024smart} & Smart Helmet: Combining Sensors, AI, Augmented Reality, and Personal Protection to Enhance First Responders' Situational Awareness & García et al. & 2024  & Object Highlighting, Augmented Enhancement, Radar \\
    \cite{gkika_object_2023} & Object detection and augmented reality annotations for increased situational awareness in light smoke conditions & Gkika et al. & 2023  & Object Highlighting \\
    \cite{grandi_design_2021} & Design and Simulation of Next-Generation Augmented Reality User Interfaces in Virtual Reality & Grandi et al. & 2021  & On Demand Menu, Augmented Enhancement, Points of Interest, Alert, 2D Map, Navigation Line, X-Ray, Object Highlighting, Annotations \\
    \cite{grandi_simulating_2020} & Simulating Next-Generation User Interfaces for Law Enforcement Traffic Stops & Grandi et al. & 2020  & Augmented Enhancement, Object Highlighting, On Demand Display, Alert \\
    \cite{arregui_augmented_2022} & An Augmented Reality Framework for First Responders: the RESPOND-A project approach & Grigoriou et al. & 2022  & Alert, Annotations, Points of Interest, 2D Map \\
    \cite{haque_augmented_2020} & Augmented reality based criminal investigation system (ARCRIME) & Haque and Saleem & 2020  & Points of Interest, Object Highlighting, Annotations \\
    \cite{ii_mixed_2020} & Mixed Reality for Post-Disaster Situational Awareness & Howard et al. & 2020  & Spatial Reconstruction \\
    \cite{hu2021human} & Human-in-the-Loop Robot-Augmented Intelligent System for Emergency Reconnaissance & Hu et al. & 2021  & Object Highlighting, Augmented Enhancement, Annotations, Points of Interest \\
    \cite{hu_seeing_2022} & Seeing through Disaster Rubble in 3D with Ground-Penetrating Radar and Interactive Augmented Reality for Urban Search and Rescue & Hu et al. & 2022  & Virtual Objects, Augmented Enhancement, Object Highlighting \\
    \cite{ierache_augmented_2016} & Augmented card system based on knowledge for medical emergency assistance & Ierache et al. & 2016  & Augmented Enhancement \\
    \cite{ismael_radiological_2023} & Radiological Incident System using Augmented Reality (RISAR) & Ismael et al. & 2023  & 2D Map, Points of Interest \\
    \cite{kapalo_comparing_2022} & Comparing Firefighters' Perceived Workload Using 2D vs. 3D Building Plans to Support Emergency Response Preplanning in a Simulated Fire Scenario & Kapolo et al. & 2020  & 3D Map \\
    \cite{karakostas_real-time_2024} & A real-time wearable AR system for egocentric vision on the edge & Karakostas et al. & 2024  & Object Highlighting, Live Video Feed, Alert \\
    \cite{kelley_guiding_2024} & Guiding Gaze: Comparing Cues for Visual Search & Kelley et al. & 2024  & Gaze Indicators \\
    \cite{kim_mobile_2016} & Mobile augmented reality in support of building damage and safety assessment & Kim et al. & 2016  & 2D Map, Radar, Spatial Reconstructuon \\
    \cite{koutitas2019virtual} & A Virtual and Augmented Reality Platform for the Training of First Responders of the Ambulance Bus & Koutitas et al. & 2019  & Points of Interest, Augmented Support \\
    \cite{lackey_new_2016} & New emergency medicine paradigm via augmented telemedicine & Kurillo et al. & 2016  & Annotations \\
    \cite{li2019fire} & A Fire Reconnaissance Robot Based on SLAM Position, Thermal Imaging Technologies, and AR Display & Li et al. & 2019  & 2D Map, Live Video Feed \\
    \cite{lugtenberg_magicbook_2023} & The MagicBook Revisited & Lugtenberg et al. & 2023  & 3D Map, Points of Interest \\
    \cite{majumdar_cloud-based_2023} & A cloud-based fire safety system for emergency responders and civic community & Majumdar et al. & 2023  & Object Highlighting, 3D Map, 2D Map, Points of Interest \\
    \cite{mannuru_mobile_2022} & Mobile AR Application for Navigation and Emergency Response & Mannuru et al. & 2022  & 2D Map, Points of Interest \\
    \cite{mantoro_pathfinding_2021} & Pathfinding for Disaster Emergency Route Using Sparse A* and Dijkstra Algorithm with Augmented Reality & Mantoro et al. & 2021  & 2D Map, Points of Interest \\
    \cite{nalamothu2024leveraging} & Leveraging Augmented Reality for Improved Situational Awareness During UAV-Driven Search and Rescue Missions & Nalamothu et al. & 2024  & Points of Interest, Augmented Enhancement \\
    \cite{nelson_user-centered_2022} & User-Centered Design and Evaluation of ARTTS: an Augmented Reality Triage Tool Suite for Mass Casualty Incidents & Nelson et al. & 2022  & Augmented Enhancement, On Demand Menu, Annotations \\
    \cite{newaz_supporting_2015} & Supporting first responder in-field communication and navigation using head-mounted displays & Newaz et al. & 2015  & 2D Map \\
    \cite{nilsson_using_2009} & Using AR to support cross-organisational collaboration in dynamic tasks & Nilsson et al. & 2009  & Points of Interest, 2D Map \\
    \cite{schmorrow_mixed-methods_2024} & A Mixed-Methods Approach for the Evaluation of Situational Awareness and User Experience with Augmented Reality Technologies & Ntoa et al. & 2024  & Object Highlighting, Live Video Feed, Alert \\
    \cite{nunes2019augmented} & An augmented reality application to support deployed emergency teams & Nunes et al. & 2018  & Augmented Compass, Points of Interest \\
    \cite{nunes_augmented_2018} & Augmented Reality in Support of Disaster Response & Nunes et al. & 2018  & Points of Interest, Compass, Radar \\
    \cite{oconnor_augmented_2023} & Augmented Reality Technology to Facilitate Proficiency in Emergency Medical Procedures & O'Connor et al. & 2023  & Augmented Support \\
    \cite{oconnor_pilot_2024} & A Pilot Randomized Controlled Trial of Augmented Reality Just-in-TimeGuidance for the Performance of Rugged Field Procedures & O'Connor et al. & 2024  & Augmented Support \\
    \cite{oregui2024augmented} & Augmented Reality Interface for Adverse-Visibility Conditions Validated by First Responders in Rescue Training Scenarios & Oregui et al. & 2024  & Radar, Augmented Enhancement, Object Highlighting \\
    \cite{pavlopoulos_augmented_1997} & An augmented reality system for health care provision via telematics support & Pavlopoulos et al. & 1997  & Annotations \\
    \cite{peretti_augmented_2022} & Augmented reality training, command and control framework for first responders & Peretti et al. & 2022  & Points of Interest, Compass, Augmented Support, 2D Map, Annotations \\
    \cite{phillips2020robotic} & A robotic augmented reality virtual window for law enforcement operations & Phillips et al. & 2020  & Object Highlighting, X-Ray \\
    \cite{balfour_next_2012} & Next generation emergency management common operating picture software/systems (COPSS) & R. E. Balfour & 2012  & Live Video Feed \\
    \cite{rainer2009simrad} & SimRad.NBC - Simulation and information system for rescue units at CBRN disasters & Rainer et al. & 2009  & Annotations, Points of Interest \\
    \cite{frasson_cpr_2023} & CPR Emergency Assistance Through Mixed Reality Communication & Rebol et al. & 2023  & Augmented Support \\
    \cite{rojas-munoz_evaluation_2020} & Evaluation of an augmented reality platform for austere surgical telementoring: a randomized controlled crossover study in cricothyroidotomies & Rojas-Muñoz et al. & 2020  & Annotations \\
    \cite{safranoglou2024augmented} & Augmented Reality for Real-time Decision-Making in Flood Emergencies & Safranoglou et al. & 2024  & Points of Interest, 3D Map, Augmented Enhancement \\
    \cite{sainidis_single-handed_2021} & Single-handed gesture UAV control and video feed AR visualization for first responders & Sainidis et al. & 2021  & Augmented Enhancement, Live Video Feed \\
    \cite{sampson_ar_2022} & AR Crew Rescue Assistant and AR Passenger Assistant Application for emergency scenarios on large passenger ships & Sampson et al. & 2022  & Points of Interest, Navigation Arrow, 2D Map \\
    \cite{schlosser_head-worn_displays_2021} & Head-Worn Displays for Emergency Medical Services Staff & Schlosser et al. & 2021  & Vitals Monitor, Annotations \\
    \cite{schlosser2024effects} & Effects of Augmented Reality-Based Remote Mentoring on Task Performance andCommunication: A Simulation Study in the Context of Emergency MedicalServices & Schlosser et al. & 2024  & Augmented Support \\
    \cite{schonauer_3d_2013} & 3D Building Reconstruction and Thermal Mapping in Fire Brigade Operations & Schönauer et al. & 2013  & 3D Map \\
    \cite{sebillo_training_2016} & Training emergency responders through augmented reality mobile interfaces & Sebillo et al. & 2016  & 2D Map, Points of Interest, Object Highlighting \\
    \cite{sharma_situational_2020} & Situational Awareness-based Augmented Reality Instructional (ARI) Module for Building Evacuation & Sharma et al. & 2020  & 2D Map, 3D Map, Points of Interest, Navigation Arrow \\
    \cite{siu_sidebars_2013} & SidebARs: Improving Awareness of off-Screen Elements in Mobile Augmented Reality & Siu and Herskovic & 2013  & Points of Interest \\
    \cite{stefanidi_real-time_2022} & Real-Time Adaptation of Context-Aware Intelligent User Interfaces, for Enhanced Situational Awareness & Stefanidi et al. & 2022  & Object Highlighting, Annotations, Icons, Alert \\
    \cite{stone_mixed_2017} & A 'mixed reality' simulator concept for future Medical Emergency Response Team training & Stone et al. & 2017  & 3D Map, Points of Interest \\
    \cite{tadokoro_robocup-rescue_2000} & RoboCup-Rescue: An international cooperative research project of robotics and AI for the disaster mitigation problem & Tadokoro et al. & 2000  & 2D Map, Points of Interest, Spatial Reconstruction \\
    \cite{tiemann_celidon_2020} & CELIDON: Supporting first responders through 3D AOA-based UWB Ad-hoclocalization & Tiemann et al. & 2020  & Points of Interest, Navigation Arrow \\
    \cite{umlauft_communication_2016} & A communication and multi-sensor solution to support dynamic generation of a situational picture & Umlauft et al. & 2016  & 2D Map, Points of Interest \\
    \cite{walker_mixed_2021} & A Mixed Reality Supervision and Telepresence Interface for Outdoor Field Robotics & Walker et al. & 2021  & 3D Map, Navigation Lines, Points of Interest \\
    \cite{wang_person--interest_2013} & Person-of-interest detection system using cloud-supported computerized-eyewear & Wang et al. & 2013  & Object Highlighting, Alert \\
    \cite{wang_method_2023} & Method and application of information sharing throughout the emergency rescue process based on 5G and AR wearable devices & Wang et al. & 2023  & Object Highlighting, Annotations \\
    \cite{weichelt_augmented_2018} & Augmented Reality Farm MAPPER Development: Lessons Learned from an App Designed to Improve Rural Emergency Response & Weichelt et al. & 2018  & Points of Interest, Compass \\
    \cite{whitlock_designing_2019} & Designing for Mobile and Immersive Visual Analytics in the Field & Whitlock et al. & 2019  & Annotations, 2D Map \\
    \cite{wilchek2025ajna} & Ajna: A Wearable Shared Perception System for Extreme Sensemaking & Wilchek et al. & 2025  & Object Highlighting, Points of Interest \\
    \cite{zhang_exploring_2024} & Exploring the Design Space of Optical See-through AR Head-Mounted Displays toSupport First Responders in the Field & Zhang et al. & 2024  & Navigation Arrow, 2D Map, 3D Map, Object Highlighting, Points of Interest, Alert \\

\end{longtable}%
\endgroup

%% file: Sections/4discussion.tex
\section{Discussion}
This section presents the findings of this scoping literature review and discusses the modes of integration, implications for research, and design considerations. Additionally, this section addresses the research questions posed in Section \ref{sec:22}. Specifically, research question two (RQ2), 'What AR UI elements have been demonstrated as effective for use in the context of public safety?', is primarily addressed throughout Section \ref{sec:analysis}. The synthesis for research question one (RQ1), 'What AR UIs have been proposed and/or used in the context of public safety?', and research question four (RQ4), 'What are the AR UI patterns used in public safety relevant for the definition of a taxonomy of such patterns?', is presented in Section \ref{sec:rq14}. Research question three (RQ3), 'What guidelines for the design and implementation of AR UIs for use in the context of public safety can be derived from the literature?', is covered in Section \ref{sec:rq3}. Additionally, the taxonomy proposed by the results of RQ4 can be found in Figure \ref{fig:taxonomy}.

\subsection{Implications for Research}\label{sec:rq14}

The findings of this review shed light on critical trends and gaps in the field of 3D user interfaces for Augmented Reality, offering new directions for advancing research. This subsection specifically synthesizes the findings for RQ1, \textit{'What AR UIs have been proposed and/or used in the context of public safety?'}, and RQ4, \textit{'What are the AR UI patterns used in public safety relevant for the definition of a taxonomy of such patterns?'}. When designing interfaces for First Responders, Navigation Interfaces were the most discussed class of interfaces, with Points of Interest (PoIs) being the most common navigation interface, representing 45.56\% of all interfaces discussed, followed by 2D maps at 32.2\%. Annotations, Object Highlighting, and Alerts were the next classes of interfaces most considered by the literature. This is due to the importance of ``Situational Awareness'' (SA) discussed by many of the papers, and how these interfaces impacted perceived and observed SA. This literature review also identified several gaps in the literature. The combination of visual feedback with other modalities such as haptic or auditory, was hardly discussed. Additionally, one paper, by Gkika et al.~\cite{gkika_object_2023}, introduced the idea of image pre-processing--a topic scarcely mentioned by other papers. Interfaces that only required audio input or no input at all by the first responders were few, however, this is an important design consideration and is discussed further in section \ref{sec:design}.

Additionally, this review identified six main functional categories under which specific AR Interface Elements (as defined in Section \ref{sec:analysis}) can be grouped. These categories are: Environment Awareness, Object Awareness, Navigation, Alerts, On-Demand Interfaces, and Augmented Guidance. While some Interface Elements showed more frequent use within particular first responder disciplines (e.g., EMS utilized more spatially augmented windows and specialized medical interfaces like Vitals Monitors, whereas firefighters often relied on HUD-based or spatial navigation elements), many Interface Elements were applied across disciplines with no significant preference.

Given the diverse application of these Interface Elements and the varying contexts, a faceted taxonomy (as presented in Figure \ref{fig:taxonomy}) was deemed most appropriate for classifying the literature. This approach allows for categorization based on multiple dimensions, such as discipline and interface requirements, rather than forcing a hierarchical structure based solely on interface type, which would obscure cross-disciplinary applications and require speculative branches for unobserved interface combinations. The faceted taxonomy is thus useful for categorizing existing research and guiding future development by allowing for the systematic consideration of different design facets. For instance, a developer could conceptualize a new AR application by selecting a specific value from each facet of the taxonomy.

\subsection{Design Considerations}\label{sec:design}\label{sec:rq3}
In response to RQ3 (\textit{'What guidelines for the design and implementation of AR UIs for use in the context of public safety can be derived from the literature?'}), this subsection discusses key design considerations. The reviewed literature suggests a need for greater generalization and modularity in interface design for first responders. While many studies presented interfaces tailored for specific tasks within a single discipline (e.g., EMS-specific tools), fewer explored designs supporting collaborative, interdisciplinary efforts or the dynamic interplay between different responder roles during an incident. For instance, the literature offers limited guidance on how a Law Enforcement Agent's AR interface might display patient vitals relayed from an EMS unit, or how EMS personnel could leverage AR for navigation in a hostile environment typically managed by firefighters. This review indicates a gap in the development of modular interfaces that can adapt or ``plug and play" information and tools relevant across different disciplines and evolving contexts. Such adaptability could enhance shared understanding and operational cohesion.

Additionally, primarily EMS literature revealed the importance of designing interfaces as hands-free as possible~\cite{nelson_user-centered_2022}. Certain papers disagreed about what degree of hands-free interaction is required. For example, Nelson et al.~\cite{nelson_user-centered_2022} who approached ``hands-free'' in the sense that hands are not occupied with a controller, while Kurillo et al.~\cite{lackey_new_2016} interpreted it to mean that the hands should not be engaged in any other task than patient care. Further research is needed to determine the most effective practices and to identify in which scenarios they are applicable.

Consideration of mental load and the operational context is another critical design guideline emerging from the literature. Several reviewed papers emphasized the need for systems that adapt interfaces to the situation, aiming to enhance situational awareness without inducing cognitive overload or stress, and using context-awareness to highlight immediate dangers~\cite{stefanidi_real-time_2022, schmorrow_mixed-methods_2024, grandi_design_2021}. These systems typically use environment and context data to augment the user's perception. The increasing focus on such context-aware systems in the reviewed literature underscores their perceived importance. However, the review also revealed that aspects such as the cognitive cost of interface navigation, particularly under stress, and rigorous evaluations of task completion times are less frequently addressed in the literature. This points to a need for more studies that not only propose AR solutions but also systematically evaluate their efficiency and compare them against traditional methods or alternative AR designs.

%% file: Sections/5conclusion.tex
\section{Conclusion}
This scoping review has systematically analyzed the landscape of Augmented Reality (AR) User Interfaces (UIs) for first responders, focusing on interface design, interaction modalities, and their application in high-stress operational environments. The review synthesized current research to identify prevalent UI elements and interaction techniques, particularly those supporting situational awareness.

A key contribution of this work is the identification of six critical facets influencing the design and application of AR UIs for first responders: operating environment, discipline, interface requirements, display hardware, display context, and output channel. Based on these facets, this review proposes a comprehensive taxonomy to categorize existing and future AR UIs in this domain. Furthermore, the analysis pinpointed 12 distinct interface elements commonly discussed in the literature: Edge Detection, Object Highlighting, X-Ray, Spatial Reconstruction, Navigation Interfaces (encompassing various elements), Alerts, On-Demand Interfaces, Gaze Indicators, Annotations, Augmented Guidance, and Vitals Monitors.

To our knowledge, this is the first scoping review to comprehensively map the domain of AR UIs for first responders. The presented taxonomy and the synthesis of interface elements provide a foundational understanding to guide future research and development, facilitating the creation of more effective and contextually appropriate AR solutions for public safety operations.

%% file: template.bbl
\begin{thebibliography}{100}
\providecommand{\url}[1]{#1}
\csname url@samestyle\endcsname
\providecommand{\newblock}{\relax}
\providecommand{\bibinfo}[2]{#2}
\providecommand{\BIBentrySTDinterwordspacing}{\spaceskip=0pt\relax}
\providecommand{\BIBentryALTinterwordstretchfactor}{4}
\providecommand{\BIBentryALTinterwordspacing}{\spaceskip=\fontdimen2\font plus
\BIBentryALTinterwordstretchfactor\fontdimen3\font minus \fontdimen4\font\relax}
\providecommand{\BIBforeignlanguage}[2]{{%
\expandafter\ifx\csname l@#1\endcsname\relax
\typeout{** WARNING: IEEEtran.bst: No hyphenation pattern has been}%
\typeout{** loaded for the language `#1'. Using the pattern for}%
\typeout{** the default language instead.}%
\else
\language=\csname l@#1\endcsname
\fi
#2}}
\providecommand{\BIBdecl}{\relax}
\BIBdecl

\bibitem{ijerph15030534}
C.~Harris, K.~McCarthy, E.~L. Liu, K.~Klein, R.~Swienton, P.~Prins, and T.~Waltz, ``Expanding understanding of response roles: An examination of immediate and first responders in the united states,'' \emph{International Journal of Environmental Research and Public Health}, vol.~15, no.~3, 2018.

\bibitem{Kedia_2022}
T.~Kedia, J.~Ratcliff, M.~O’Connor, S.~Oluic, M.~Rose, J.~Freeman, and K.~Rainwater-Lovett, ``Technologies enabling situational awareness during disaster response: A systematic review,'' \emph{Disaster Medicine and Public Health Preparedness}, vol.~16, no.~1, p. 341–359, 2022.

\bibitem{kitchenham2007guidelines}
S.~C. B.~Kitchenham, ``Guidelines for performing systematic literature reviews in software engineering,'' Technical report, ver. 2.3 ebse technical report. ebse, Tech. Rep., 2007.

\bibitem{petticrew2008systematic}
M.~Petticrew and H.~Roberts, \emph{Systematic reviews in the social sciences: A practical guide}.\hskip 1em plus 0.5em minus 0.4em\relax John Wiley \& Sons, 2008.

\bibitem{arksey2005scoping}
H.~Arksey and L.~O’Malley, ``Scoping studies: towards a methodological framework,'' \emph{International Journal of Social Research Methodology}, vol.~8, no.~1, pp. 19--32, 2005.

\bibitem{mays2001systematic}
N.~Mays, E.~Roberts, and J.~Popay, ``Systematic reviews: synthesis of best evidence for health care decisions,'' \emph{BMJ}, vol. 323, no. 7312, pp. 765--768, 2001.

\bibitem{peters2015guidance}
M.~D. Peters, C.~M. Godfrey, H.~Khalil, P.~McInerney, D.~Parker, and C.~Baldini~Soares, ``Guidance for conducting systematic scoping reviews,'' \emph{International Journal of Evidence-Based Healthcare}, vol.~13, no.~3, pp. 141--146, 2015.

\bibitem{parsifal}
{Parsifal}, ``{Parsifal}: Perform systematic literature reviews,'' \url{https://parsif.al/}, n.d., accessed: 2025-05-30.

\bibitem{antevski_5g-based_2021}
K.~Antevski, L.~Girletti, C.~J. Bernardos, A.~De~La~Oliva, J.~Baranda, and J.~Mangues-Bafalluy, ``\BIBforeignlanguage{en}{A {5G}-{Based} {eHealth} {Monitoring} and {Emergency} {Response} {System}: {Experience} and {Lessons} {Learned}},'' \emph{\BIBforeignlanguage{en}{IEEE Access}}, vol.~9, pp. 131\,420--131\,429, 2021.

\bibitem{azpiroz2024white}
I.~Azpiroz, I.~G. Olaizola, X.~Oregui, A.~F. Garc{\'\i}a, V.~Ruiz, B.~Larraga-Garc{\'\i}a, and {\'A}.~Guti{\'e}rrez, ``White paper on adaptive situational awareness enhancing augmented reality interface design on first responders in rescue tasks,'' \emph{Applied Sciences}, vol.~14, no.~18, p. 8282, 2024.

\bibitem{schmorrow_implementing_2016}
T.~Bailie, J.~Martin, Z.~Aman, R.~Brill, and A.~Herman, ``\BIBforeignlanguage{en}{Implementing {User}-{Centered} {Methods} and {Virtual} {Reality} to {Rapidly} {Prototype} {Augmented} {Reality} {Tools} for {Firefighters}},'' in \emph{\BIBforeignlanguage{en}{Foundations of {Augmented} {Cognition}: {Neuroergonomics} and {Operational} {Neuroscience}}}, D.~D. Schmorrow and C.~M. Fidopiastis, Eds.\hskip 1em plus 0.5em minus 0.4em\relax Cham: Springer International Publishing, 2016, vol. 9744, pp. 135--144, series Title: Lecture Notes in Computer Science.

\bibitem{berndt_optical_2015}
H.~Berndt, T.~Mentler, and M.~Herczeg, ``\BIBforeignlanguage{en}{Optical {Head}-{Mounted} {Displays} in {Mass} {Casualty} {Incidents}: {Keeping} an {Eye} on {Patients} and {Hazardous} {Materials}},'' \emph{\BIBforeignlanguage{en}{International Journal of Information Systems for Crisis Response and Management}}, vol.~7, no.~3, pp. 1--15, Jul. 2015.

\bibitem{bhattarai_embedded_2020}
M.~Bhattarai, A.~R. Jensen-Curtis, and M.~Martinez-Ramon, ``\BIBforeignlanguage{en}{An embedded deep learning system for augmented reality in firefighting applications},'' in \emph{\BIBforeignlanguage{en}{2020 19th {IEEE} {International} {Conference} on {Machine} {Learning} and {Applications} ({ICMLA})}}.\hskip 1em plus 0.5em minus 0.4em\relax Miami, FL, USA: IEEE, Dec. 2020, pp. 1224--1230.

\bibitem{bram-larbi_collision_2020}
K.~Bram-Larbi, V.~Charissis, S.~Khan, R.~Lagoo, D.~K. Harrison, and D.~Drikakis, ``\BIBforeignlanguage{en}{Collision {Avoidance} {Head}-{Up} {Display}: {Design} {Considerations} for {Emergency} {Services}’ {Vehicles}},'' in \emph{\BIBforeignlanguage{en}{2020 {IEEE} {International} {Conference} on {Consumer} {Electronics} ({ICCE})}}.\hskip 1em plus 0.5em minus 0.4em\relax Las Vegas, NV, USA: IEEE, Jan. 2020, pp. 1--7.

\bibitem{brandao_using_2017}
W.~L. Brandao and M.~S. Pinho, ``\BIBforeignlanguage{en}{Using augmented reality to improve dismounted operators' situation awareness},'' in \emph{\BIBforeignlanguage{en}{2017 {IEEE} {Virtual} {Reality} ({VR})}}.\hskip 1em plus 0.5em minus 0.4em\relax Los Angeles, CA, USA: IEEE, 2017, pp. 297--298.

\bibitem{broach_usability_2018}
J.~Broach, A.~Hart, M.~Griswold, J.~Lai, E.~W. Boyer, A.~B. Skolnik, and P.~R. Chai, ``Usability and reliability of smart glasses for secondary triage during mass casualty incidents,'' in \emph{Proceedings of the Annual Hawaii International Conference on System Sciences (HICSS)}, 2018, pp. 1416--1422.

\bibitem{cai_mobile_2023}
Y.~Cai and M.~Siegel, ``\BIBforeignlanguage{en}{Mobile incident command dashboard ({MIC}-{D})},'' \emph{\BIBforeignlanguage{en}{Electronic Imaging}}, vol.~35, no.~3, pp. 358--1--358--4, Jan. 2023.

\bibitem{campos_mobile_2019}
A.~Campos, N.~Correia, T.~Romão, I.~Nunes, and M.~Simões-Marques, ``\BIBforeignlanguage{en}{Mobile augmented reality techniques for emergency response},'' in \emph{\BIBforeignlanguage{en}{Proceedings of the 16th {EAI} {International} {Conference} on {Mobile} and {Ubiquitous} {Systems}: {Computing}, {Networking} and {Services}}}.\hskip 1em plus 0.5em minus 0.4em\relax Houston Texas USA: ACM, Nov. 2019, pp. 31--39.

\bibitem{chae_design_2023}
Y.-J. Chae, H.-W. Lee, J.-H. Kim, S.-W. Hwang, and Y.-Y. Park, ``\BIBforeignlanguage{en}{Design of a {Mixed} {Reality} {System} for {Simulating} {Indoor} {Disaster} {Rescue}},'' \emph{\BIBforeignlanguage{en}{Applied Sciences}}, vol.~13, no.~7, p. 4418, Mar. 2023.

\bibitem{chalimas_cross-device_2023}
T.~Chalimas and K.~Mania, ``\BIBforeignlanguage{en}{Cross-{Device} {Augmented} {Reality} {Systems} for {Fire} and {Rescue} based on {Thermal} {Imaging} and {Live} {Tracking}},'' in \emph{\BIBforeignlanguage{en}{2023 {IEEE} {International} {Symposium} on {Mixed} and {Augmented} {Reality} {Adjunct} ({ISMAR}-{Adjunct})}}.\hskip 1em plus 0.5em minus 0.4em\relax Sydney, Australia: IEEE, Oct. 2023, pp. 50--54.

\bibitem{chan_erwear_2016}
E.~Chan, Y.~Wang, T.~Seyed, and F.~Maurer, ``\BIBforeignlanguage{en}{{ERWear}: {Wearable} {System} {Design} through the {Lens} of {First} {Responders}},'' in \emph{\BIBforeignlanguage{en}{Proceedings of the 2016 {ACM} {International} {Conference} on {Interactive} {Surfaces} and {Spaces}}}.\hskip 1em plus 0.5em minus 0.4em\relax Niagara Falls Ontario Canada: ACM, Nov. 2016, pp. 489--492.

\bibitem{cheng_effective_2023}
Z.~Cheng, G.~Ren, S.~Dong, R.~Li, Z.~Huo, J.~Li, and G.~Wang, ``\BIBforeignlanguage{en}{Effective {Navigation} for an {Emergency} {Response} for {Firefighters} with {Wearable} {Haptic} {Displays} {Design}},'' in \emph{\BIBforeignlanguage{en}{2023 {IEEE} 6th {International} {Conference} on {Knowledge} {Innovation} and {Invention} ({ICKII})}}.\hskip 1em plus 0.5em minus 0.4em\relax Sapporo, Japan: IEEE, Aug. 2023, pp. 46--49.

\bibitem{christaki_augmented_2022}
K.~Christaki, D.~Tsiakmakis, I.~Babic, G.~Inglese, K.~Konstantoudakis, G.~Giunta, A.~Dimou, O.~Balet, and P.~Daras, ``Augmented reality points of interest for improved first responder situational awareness,'' in \emph{Proceedings of the 19th International Conference on Information Systems for Crisis Response and Management (ISCRAM)}, Aix-en-Provence, France, 2022, pp. 755--770.

\bibitem{christian_google_2014}
M.~Christian, A.~Depaz, M.~Grimm, J.~W. Lartigue, R.~Sweatland, and C.~Talley, ``\BIBforeignlanguage{en}{Google glass for public safety: leveraging {Google} {Glass} for automatic information retrieval and notification by public safety officers in the field},'' in \emph{\BIBforeignlanguage{en}{Proceedings of the 2014 {ACM} {Southeast} {Regional} {Conference}}}.\hskip 1em plus 0.5em minus 0.4em\relax Kennesaw Georgia: ACM, Mar. 2014, pp. 1--3.

\bibitem{chroust2009training}
G.~Chroust, S.~Sch{\"o}nhacker, K.~Rainer, M.~Roth, and P.~Ziehesberger, ``Training and supporting first responders by mixed reality environments,'' in \emph{Proceedings of the 53rd Annual Meeting of the ISSS-2009, Brisbane, Australia}, 2009.

\bibitem{crowley_ar_2014}
D.~E. Crowley, R.~R. Murphy, A.~McNamara, T.~D. McLaughlin, and B.~A. Duncan, ``\BIBforeignlanguage{en}{{AR} browser for points of interest in disaster response in {UAV} imagery},'' in \emph{\BIBforeignlanguage{en}{{CHI} '14 {Extended} {Abstracts} on {Human} {Factors} in {Computing} {Systems}}}.\hskip 1em plus 0.5em minus 0.4em\relax Toronto Ontario Canada: ACM, Apr. 2014, pp. 2173--2178.

\bibitem{dave_augmenting_2013}
R.~Dave, S.~K. Boddhu, M.~McCartney, and J.~West, ``\BIBforeignlanguage{en}{Augmenting {Situational} {Awareness} for {First} responders using {Social} media as a sensor},'' \emph{\BIBforeignlanguage{en}{IFAC Proceedings Volumes}}, vol.~46, no.~15, pp. 133--140, 2013.

\bibitem{davis2002augmented}
L.~Davis, Y.~Ha, S.~Frolich, G.~Martin, C.~Meyer, B.~Pettitt, J.~Norfleet, K.-C. Lin, and J.~P. Rolland, ``Augmented reality and training for airway management procedures,'' in \emph{Medicine Meets Virtual Reality 02/10}.\hskip 1em plus 0.5em minus 0.4em\relax IOS Press, 2002, pp. 121--126.

\bibitem{demirkan_underground_2024}
D.~C. Demirkan, A.~Segal, A.~Mallik, S.~Duzgun, and A.~J. Petruska, ``\BIBforeignlanguage{en}{Underground mine emergency evacuation planning: {AR} implementation and case study},'' in \emph{\BIBforeignlanguage{en}{Proceendings of Optical Architectures for Displays and Sensing in Augmented, Virtual, and Mixed Reality {(AR, VR, MR)}}}, H.~Hua, N.~Argaman, and D.~K. Nikolov, Eds., International Society for Optics and Photonics.\hskip 1em plus 0.5em minus 0.4em\relax San Francisco, United States: SPIE, Mar. 2024.

\bibitem{dianxi_design_2021}
Z.~Dianxi, C.~Danhongand, and G.~Zhen, ``\BIBforeignlanguage{en}{Design of {Intelligent} {Firefighter} {Helmet} {Based} on {AR} {Technology}},'' \emph{\BIBforeignlanguage{en}{IOP Conference Series: Earth and Environmental Science}}, vol. 621, no.~1, p. 012176, Jan. 2021.

\bibitem{engelbrecht_viability_2018}
H.~Engelbrecht and S.~G. Lukosch, ``\BIBforeignlanguage{en}{Viability of {Augmented} {Content} for {Field} {Policing}},'' in \emph{\BIBforeignlanguage{en}{2018 {IEEE} {International} {Symposium} on {Mixed} and {Augmented} {Reality} {Adjunct} ({ISMAR}-{Adjunct})}}.\hskip 1em plus 0.5em minus 0.4em\relax Munich, Germany: IEEE, Oct. 2018, pp. 386--389.

\bibitem{del2014red}
M.~del Roc{\'\i}o Fuentes~Fern{\'a}ndez, C.~I.~T. Bernabe, and R.~R. Rodr{\'\i}guez, ``Red cross triage app design for augmented reality glasses,'' in \emph{Proceedings of the 5th Mexican Conference on Human-Computer Interaction}, 2014, pp. 11--14.

\bibitem{follmann2019technical}
A.~Follmann, M.~Ohligs, N.~Hochhausen, S.~K. Beckers, R.~Rossaint, and M.~Czaplik, ``Technical support by smart glasses during a mass casualty incident: a randomized controlled simulation trial on technically assisted triage and telemedical app use in disaster medicine,'' \emph{Journal of medical Internet research}, vol.~21, no.~1, p. e11939, 2019.

\bibitem{garcia_health-5g_2023}
F.~M. García, R.~Moraleda, S.~Schez-Sobrino, D.~N. Monekosso, D.~Vallejo, and C.~Glez-Morcillo, ``\BIBforeignlanguage{en}{Health-{5G}: {A} {Mixed} {Reality}-{Based} {System} for {Remote} {Medical} {Assistance} in {Emergency} {Situations}},'' \emph{\BIBforeignlanguage{en}{IEEE Access}}, vol.~11, pp. 59\,016--59\,032, 2023.

\bibitem{garcia2024smart}
A.~F. García, X.~O. Biain, K.~Lingos, K.~Konstantoudakis, A.~B. Hernández, I.~A. Iragorri, and D.~Zarpalas, ``Smart helmet: Combining sensors, {AI}, augmented reality, and personal protection to enhance first responders’ situational awareness,'' \emph{IT Professional}, vol.~25, no.~6, pp. 45--53, 2023.

\bibitem{gkika_object_2023}
I.~Gkika, D.~Pattas, K.~Konstantoudakis, and D.~Zarpalas, ``Object detection and augmented reality annotations for increased situational awareness in light smoke conditions,'' in \emph{Proceedings of the 20th International Conference on Information Systems for Crisis Response and Management (ISCRAM)}, Omaha, USA, 2023, pp. 231--241.

\bibitem{grandi_design_2021}
J.~G. Grandi, Z.~Cao, M.~Ogren, and R.~Kopper, ``\BIBforeignlanguage{en}{Design and {Simulation} of {Next}-{Generation} {Augmented} {Reality} {User} {Interfaces} in {Virtual} {Reality}},'' in \emph{\BIBforeignlanguage{en}{2021 {IEEE} {Conference} on {Virtual} {Reality} and {3D} {User} {Interfaces} {Abstracts} and {Workshops} ({VRW})}}.\hskip 1em plus 0.5em minus 0.4em\relax Lisbon, Portugal: IEEE, Mar. 2021, pp. 23--29.

\bibitem{grandi_simulating_2020}
------, ``\BIBforeignlanguage{en}{Simulating {Next}-{Generation} {User} {Interfaces} for {Law} {Enforcement} {Traffic} {Stops}},'' in \emph{\BIBforeignlanguage{en}{2020 {IEEE} {Conference} on {Virtual} {Reality} and {3D} {User} {Interfaces} {Abstracts} and {Workshops} ({VRW})}}.\hskip 1em plus 0.5em minus 0.4em\relax Atlanta, GA, USA: IEEE, Mar. 2020, pp. 826--827.

\bibitem{arregui_augmented_2022}
H.~Arregui, E.~Irigoyen, I.~Cejudo, S.~Simonsen, D.~Ribar, M.-A. Kourtis, Y.~Spyridis, N.~Stathakarou, and M.~C. Batistatos, ``\BIBforeignlanguage{en}{An {Augmented} {Reality} {Framework} for {First} {Responders}: the {RESPOND}-{A} project approach},'' in \emph{\BIBforeignlanguage{en}{2022 {Panhellenic} {Conference} on {Electronics} \& {Telecommunications} ({PACET})}}.\hskip 1em plus 0.5em minus 0.4em\relax Tripolis, Greece: IEEE, Dec. 2022, pp. 1--6.

\bibitem{haque_augmented_2020}
S.~E.~I. Haque and S.~Saleem, ``\BIBforeignlanguage{en}{Augmented reality based criminal investigation system ({ARCRIME})},'' in \emph{\BIBforeignlanguage{en}{2020 8th {International} {Symposium} on {Digital} {Forensics} and {Security} ({ISDFS})}}.\hskip 1em plus 0.5em minus 0.4em\relax Beirut, Lebanon: IEEE, Jun. 2020, pp. 1--6.

\bibitem{ii_mixed_2020}
J.~P. Howard~II, A.~O. Tucker~IV, S.~A. Bailey, and J.~L. Dean, ``\BIBforeignlanguage{en}{Mixed {Reality} for {Post}-{Disaster} {Situational} {Awareness}},'' \emph{\BIBforeignlanguage{en}{Johns Hopkins APL Technical Digest}}, vol.~35, no.~3, 2020.

\bibitem{hu2021human}
D.~Hu, S.~Li, J.~Du, and J.~Cai, ``Human-in-the-loop robot-augmented intelligent system for emergency reconnaissance,'' in \emph{Computing in Civil Engineering 2021}, 2021, pp. 1409--1416.

\bibitem{hu_seeing_2022}
D.~Hu, L.~Chen, J.~Du, J.~Cai, and S.~Li, ``Seeing through disaster rubble in 3d with ground‐penetrating radar and interactive augmented reality for urban search and rescue,'' \emph{Journal of Computing in Civil Engineering}, vol.~36, no.~5, p. 04022021, 2022.

\bibitem{ierache_augmented_2016}
J.~Ierache, N.~Mangiarua, N.~Verdicchio, D.~Sanz, C.~Montalvo, F.~Petrolo, and S.~Igarza, ``\BIBforeignlanguage{en}{Augmented card system based on knowledge for medical emergency assistance},'' in \emph{\BIBforeignlanguage{en}{2016 {IEEE} {Congreso} {Argentino} de {Ciencias} de la {Informática} y {Desarrollos} de {Investigación} ({CACIDI})}}.\hskip 1em plus 0.5em minus 0.4em\relax Buenos Aires, Argentina: IEEE, Nov. 2016, pp. 1--3.

\bibitem{ismael_radiological_2023}
M.~Ismael, R.~McCall, M.~Cornil, M.~Griffin, and J.~Baixauli, ``\BIBforeignlanguage{en}{Radiological {Incident} {System} using {Augmented} {Reality} ({RISAR})},'' in \emph{\BIBforeignlanguage{en}{2023 {IEEE} {Conference} on {Virtual} {Reality} and {3D} {User} {Interfaces} {Abstracts} and {Workshops} ({VRW})}}.\hskip 1em plus 0.5em minus 0.4em\relax Shanghai, China: IEEE, Mar. 2023, pp. 591--592.

\bibitem{kapalo_comparing_2022}
K.~A. Kapalo, K.~Pfeil, J.~Bonnell, and J.~LaViola, ``\BIBforeignlanguage{en}{Comparing {Firefighters}' {Perceived} {Workload} {Using} {2D} vs. {3D} {Building} {Plans} to {Support} {Emergency} {Response} {Preplanning} in a {Simulated} {Fire} {Scenario}},'' in \emph{\BIBforeignlanguage{en}{2022 {IEEE} {International} {Symposium} on {Mixed} and {Augmented} {Reality} {Adjunct} ({ISMAR}-{Adjunct})}}.\hskip 1em plus 0.5em minus 0.4em\relax Singapore, Singapore: IEEE, Oct. 2022, pp. 634--639.

\bibitem{karakostas_real-time_2024}
I.~Karakostas, A.~Valakou, D.~Gavgiotaki, Z.~Stefanidi, I.~Pastaltzidis, G.~Tsipouridis, N.~Kilis, K.~C. Apostolakis, S.~Ntoa, N.~Dimitriou, G.~Margetis, and D.~Tzovaras, ``\BIBforeignlanguage{en}{A real-time wearable {AR} system for egocentric vision on the edge},'' \emph{\BIBforeignlanguage{en}{Virtual Reality}}, vol.~28, no.~1, p.~44, Mar. 2024.

\bibitem{kelley_guiding_2024}
B.~Kelley, C.~Wickens, B.~Clegg, A.~C. Warden, and F.~R. Ortega, ``\BIBforeignlanguage{en}{Guiding {Gaze}: {Comparing} {Cues} for {Visual} {Search}},'' in \emph{\BIBforeignlanguage{en}{2024 {IEEE} {Conference} on {Virtual} {Reality} and {3D} {User} {Interfaces} {Abstracts} and {Workshops} ({VRW})}}.\hskip 1em plus 0.5em minus 0.4em\relax Orlando, FL, USA: IEEE, Mar. 2024, pp. 1053--1054.

\bibitem{kim_mobile_2016}
W.~Kim, N.~Kerle, and M.~Gerke, ``\BIBforeignlanguage{en}{Mobile augmented reality in support of building damage and safety assessment},'' \emph{\BIBforeignlanguage{en}{Natural Hazards and Earth System Sciences}}, vol.~16, no.~1, pp. 287--298, Feb. 2016.

\bibitem{lackey_new_2016}
G.~Kurillo, A.~Y. Yang, V.~Shia, A.~Bair, and R.~Bajcsy, ``\BIBforeignlanguage{en}{New emergency medicine paradigm via augmented telemedicine},'' in \emph{\BIBforeignlanguage{en}{Virtual, Augmented and Mixed Reality}}, S.~Lackey and R.~Shumaker, Eds.\hskip 1em plus 0.5em minus 0.4em\relax Cham: Springer International Publishing, 2016, vol. 9740, pp. 502--511, series Title: Lecture Notes in Computer Science.

\bibitem{li2019fire}
S.~Li, C.~Feng, Y.~Niu, L.~Shi, Z.~Wu, and H.~Song, ``A fire reconnaissance robot based on slam position, thermal imaging technologies, and ar display,'' \emph{Sensors}, vol.~19, no.~22, p. 5036, 2019.

\bibitem{mannuru_mobile_2022}
N.~R. Mannuru, M.~Kanumuru, and S.~Sharma, ``Mobile {AR} {Application} for {Navigation} and {Emergency} {Response},'' in \emph{2022 {International} {Conference} on {Computational} {Science} and {Computational} {Intelligence} ({CSCI})}.\hskip 1em plus 0.5em minus 0.4em\relax Las Vegas, NV, USA: IEEE, Dec. 2022, pp. 1137--1142.

\bibitem{mantoro_pathfinding_2021}
T.~Mantoro, Z.~Alamsyah, and M.~A. Ayu, ``\BIBforeignlanguage{en}{Pathfinding for {Disaster} {Emergency} {Route} {Using} {Sparse} {A}* and {Dijkstra} {Algorithm} with {Augmented} {Reality}},'' in \emph{\BIBforeignlanguage{en}{2021 {IEEE} 7th {International} {Conference} on {Computing}, {Engineering} and {Design} ({ICCED})}}.\hskip 1em plus 0.5em minus 0.4em\relax Sukabumi, Indonesia: IEEE, Aug. 2021, pp. 1--6.

\bibitem{nalamothu2024leveraging}
R.~Nalamothu, P.~Sontha, J.~Karravula, and A.~Agrawal, ``Leveraging augmented reality for improved situational awareness during uav-driven search and rescue missions,'' in \emph{2024 IEEE International Symposium on Safety Security Rescue Robotics (SSRR)}.\hskip 1em plus 0.5em minus 0.4em\relax IEEE, 2024, pp. 221--228.

\bibitem{nelson_user-centered_2022}
C.~R. Nelson, J.~L. Gabbard, J.~B. Moats, and R.~K. Mehta, ``\BIBforeignlanguage{en}{User-{Centered} {Design} and {Evaluation} of {ARTTS}: an {Augmented} {Reality} {Triage} {Tool} {Suite} for {Mass} {Casualty} {Incidents}},'' in \emph{\BIBforeignlanguage{en}{2022 {IEEE} {International} {Symposium} on {Mixed} and {Augmented} {Reality} ({ISMAR})}}.\hskip 1em plus 0.5em minus 0.4em\relax Singapore, Singapore: IEEE, Oct. 2022, pp. 336--345.

\bibitem{newaz_supporting_2015}
F.~B.~M. Newaz, A.~W. Eide, and A.~Pultier, ``Supporting first responder in-field communication and navigation using head-mounted displays,'' in \emph{Proceedings of the 12th International Conference on Information Systems for Crisis Response and Management (ISCRAM)}, Kristiansand, Norway, 2015.

\bibitem{schmorrow_mixed-methods_2024}
S.~Ntoa, G.~Margetis, A.~Valakou, F.~Makri, N.~Dimitriou, I.~Karakostas, G.~Kokkinis, K.~C. Apostolakis, D.~Tzovaras, and C.~Stephanidis, ``\BIBforeignlanguage{en}{A {Mixed}-{Methods} {Approach} for the {Evaluation} of {Situational} {Awareness} and {User} {Experience} with {Augmented} {Reality} {Technologies}},'' in \emph{\BIBforeignlanguage{en}{Augmented {Cognition}}}, D.~D. Schmorrow and C.~M. Fidopiastis, Eds.\hskip 1em plus 0.5em minus 0.4em\relax Cham: Springer Nature Switzerland, 2024, vol. 14694, pp. 199--219, series Title: Lecture Notes in Computer Science.

\bibitem{nunes2019augmented}
I.~L. Nunes, R.~Lucas, M.~Sim{\~o}es-Marques, and N.~Correia, ``An augmented reality application to support deployed emergency teams,'' in \emph{Proceedings of the 20th Congress of the International Ergonomics Association (IEA 2018) Volume V: Human Simulation and Virtual Environments, Work With Computing Systems (WWCS), Process Control 20}.\hskip 1em plus 0.5em minus 0.4em\relax Springer, 2019, pp. 195--204.

\bibitem{nunes_augmented_2018}
I.~L. Nunes, R.~Lucas, M.~Simões-Marques, and N.~Correia, ``\BIBforeignlanguage{en}{Augmented {Reality} in {Support} of {Disaster} {Response}},'' in \emph{\BIBforeignlanguage{en}{Advances in {Human} {Factors} and {Systems} {Interaction}}}, I.~L. Nunes, Ed.\hskip 1em plus 0.5em minus 0.4em\relax Cham: Springer International Publishing, 2018, vol. 592, pp. 155--167, series Title: Advances in Intelligent Systems and Computing.

\bibitem{oconnor_augmented_2023}
L.~O’Connor, S.~Zamani, L.~Porter, N.~McGeorge, S.~Latiff, T.~Boardman, M.~Loconte, M.~Weiner, E.~McGarry, F.~Pina, J.~A. Hermaan, A.~Milsten, M.~Reznek, and J.~Broach, ``Augmented reality technology to facilitate proficiency in emergency medical procedures,'' in \emph{Proceedings of the Annual Hawaii International Conference on System Sciences (HICSS)}, 2023, pp. 3132--3140.

\bibitem{oconnor_pilot_2024}
L.~O’Connor, S.~Zamani, X.~Ding, N.~McGeorge, S.~Latiff, C.~Liu, J.~Acevedo~Herman, M.~LoConte, A.~Milsten, M.~Weiner, T.~Boardman, M.~Reznek, M.~Hall, and J.~P. Broach, ``\BIBforeignlanguage{en}{A {Pilot} {Randomized} {Controlled} {Trial} of {Augmented} {Reality} {Just}-in-{Time} {Guidance} for the {Performance} of {Rugged} {Field} {Procedures}},'' \emph{\BIBforeignlanguage{en}{Prehospital and Disaster Medicine}}, pp. 1--9, May 2024.

\bibitem{oregui2024augmented}
X.~Oregui, A.~Fern{\'a}ndez~Garc{\'\i}a, I.~Azpiroz, B.~Larraga-Garc{\'\i}a, V.~Ruiz, I.~Garc{\'\i}a~Olaizola, and {\'A}.~Guti{\'e}rrez, ``Augmented reality interface for adverse-visibility conditions validated by first responders in rescue training scenarios,'' \emph{Electronics}, vol.~13, no.~18, p. 3739, 2024.

\bibitem{pavlopoulos_augmented_1997}
S.~Pavlopoulos, S.~Dembeyiotis, and D.~Koutsouris, ``An augmented reality system for health care provision via telematics support,'' in \emph{Technology and Informatics: Medicine Meets Virtual Reality}, ser. Studies in Health Technology and Informatics, K.~S. Morgan, Ed.\hskip 1em plus 0.5em minus 0.4em\relax Amsterdam, The Netherlands: IOS Press, 1997, vol.~39, pp. 286--288.

\bibitem{phillips2020robotic}
N.~Phillips, B.~Kruse, F.~A. Khan, J.~E. Swan~II, and C.~L. Bethel, ``A robotic augmented reality virtual window for law enforcement operations,'' in \emph{Virtual, Augmented and Mixed Reality. Design and Interaction: 12th International Conference, VAMR 2020, Held as Part of the 22nd HCI International Conference, HCII 2020, Copenhagen, Denmark, July 19--24, 2020, Proceedings, Part I 22}.\hskip 1em plus 0.5em minus 0.4em\relax Springer, 2020, pp. 591--610.

\bibitem{balfour_next_2012}
R.~E. Balfour, ``\BIBforeignlanguage{en}{Next generation emergency management common operating picture software/systems ({COPSS})},'' in \emph{\BIBforeignlanguage{en}{2012 {IEEE} {Long} {Island} {Systems}, {Applications} and {Technology} {Conference} ({LISAT})}}.\hskip 1em plus 0.5em minus 0.4em\relax Farmingdale, NY, USA: IEEE, May 2012, pp. 1--6.

\bibitem{rainer2009simrad}
K.~Rainer, N.~Sturm, S.~Sch{\"o}nhacker, and G.~Chroust, ``Simrad. nbc--simulation and information system for rescue units at cbrn disasters,'' in \emph{Intelligent Distributed Computing III: Proceedings of the 3rd International Symposium on Intelligent Distributed Computing--IDC 2009, Ayia Napa, Cyprus, October 2009}.\hskip 1em plus 0.5em minus 0.4em\relax Springer, 2009, pp. 297--303.

\bibitem{frasson_cpr_2023}
M.~Rebol, A.~Steinmaurer, F.~Gamillscheg, K.~Pietroszek, C.~Gütl, C.~Ranniger, C.~Hood, A.~Rutenberg, and N.~Sikka, ``\BIBforeignlanguage{en}{{CPR} {Emergency} {Assistance} {Through} {Mixed} {Reality} {Communication}},'' in \emph{\BIBforeignlanguage{en}{Augmented {Intelligence} and {Intelligent} {Tutoring} {Systems}}}, C.~Frasson, P.~Mylonas, and C.~Troussas, Eds.\hskip 1em plus 0.5em minus 0.4em\relax Cham: Springer Nature Switzerland, 2023, vol. 13891, pp. 415--429, series Title: Lecture Notes in Computer Science.

\bibitem{rojas-munoz_evaluation_2020}
E.~Rojas-Muñoz, C.~Lin, N.~Sanchez-Tamayo, M.~E. Cabrera, D.~Andersen, V.~Popescu, J.~A. Barragan, B.~Zarzaur, P.~Murphy, K.~Anderson, T.~Douglas, C.~Griffis, J.~McKee, A.~W. Kirkpatrick, and J.~P. Wachs, ``\BIBforeignlanguage{en}{Evaluation of an augmented reality platform for austere surgical telementoring: a randomized controlled crossover study in cricothyroidotomies},'' \emph{\BIBforeignlanguage{en}{npj Digital Medicine}}, vol.~3, no.~1, p.~75, May 2020.

\bibitem{safranoglou2024augmented}
I.~Safranoglou, A.~Stavroulakis, M.~Ebel, J.~Pottebaum, G.~Lamprinakis, D.~Dimelli, and K.~Mania, ``Augmented reality for real-time decision-making in flood emergencies,'' in \emph{2024 IEEE International Symposium on Mixed and Augmented Reality Adjunct (ISMAR-Adjunct)}.\hskip 1em plus 0.5em minus 0.4em\relax IEEE, 2024, pp. 110--116.

\bibitem{sainidis_single-handed_2021}
D.~Sainidis, D.~Tsiakmakis, K.~Konstantoudakis, G.~Albanis, A.~Dimou, and P.~Daras, ``Single-handed gesture uav control and video feed ar visualization for first responders,'' in \emph{Proceedings of the 18th International Conference on Information Systems for Crisis Response and Management (ISCRAM)}, Blacksburg, VA, USA, 2021, pp. 835--848.

\bibitem{sampson_ar_2022}
O.~Sampson, S.~Bolierakis, M.~Krommyda, L.~Karagianidis, and A.~Amditis, ``\BIBforeignlanguage{en}{{AR} {Crew} {Rescue} {Assistant} and {AR} {Passenger} {Assistant} {Application} for emergency scenarios on large passenger ships},'' in \emph{\BIBforeignlanguage{en}{2022 {IEEE} {International} {Conference} on {Imaging} {Systems} and {Techniques} ({IST})}}.\hskip 1em plus 0.5em minus 0.4em\relax Kaohsiung, Taiwan: IEEE, Jun. 2022, pp. 1--6.

\bibitem{schlosser_head-worn_displays_2021}
P.~Schlosser, B.~Matthews, I.~Salisbury, P.~Sanderson, and S.~Hayes, ``\BIBforeignlanguage{en}{Head-{Worn} {Displays} for {Emergency} {Medical} {Services} {Staff}: {Properties} of {Prehospital} {Work}, {Use} {Cases}, and {Design} {Considerations}},'' in \emph{\BIBforeignlanguage{en}{Proceedings of the 2021 {CHI} {Conference} on {Human} {Factors} in {Computing} {Systems}}}.\hskip 1em plus 0.5em minus 0.4em\relax Yokohama Japan: ACM, May 2021, pp. 1--14.

\bibitem{schlosser2024effects}
P.~D. Schlosser, B.~Matthews, P.~M. Sanderson, A.~Donohue, and S.~Hayes, ``Effects of augmented reality-based remote mentoring on task performance and communication: A simulation study in the context of emergency medical services,'' \emph{Telemedicine and e-Health}, vol.~30, no.~5, pp. 1470--1478, 2024.

\bibitem{schonauer_3d_2013}
C.~Sch{\"o}nauer, E.~Vonach, G.~Gerstweiler, and H.~Kaufmann, ``3d building reconstruction and thermal mapping in fire brigade operations,'' in \emph{2013 IEEE Virtual Reality (VR)}, 2013.

\bibitem{sebillo_training_2016}
M.~Sebillo, G.~Vitiello, L.~Paolino, and A.~Ginige, ``\BIBforeignlanguage{en}{Training emergency responders through augmented reality mobile interfaces},'' \emph{\BIBforeignlanguage{en}{Multimedia Tools and Applications}}, vol.~75, no.~16, pp. 9609--9622, Aug. 2016.

\bibitem{sharma_situational_2020}
S.~Sharma, J.~Stigall, and S.~T. Bodempudi, ``\BIBforeignlanguage{en}{Situational {Awareness}-based {Augmented} {Reality} {Instructional} ({ARI}) {Module} for {Building} {Evacuation}},'' in \emph{\BIBforeignlanguage{en}{2020 {IEEE} {Conference} on {Virtual} {Reality} and {3D} {User} {Interfaces} {Abstracts} and {Workshops} ({VRW})}}.\hskip 1em plus 0.5em minus 0.4em\relax Atlanta, GA, USA: IEEE, Mar. 2020, pp. 70--78.

\bibitem{siu_sidebars_2013}
T.~Siu and V.~Herskovic, ``\BIBforeignlanguage{en}{{SidebARs}: improving awareness of off-screen elements in mobile augmented reality},'' in \emph{\BIBforeignlanguage{en}{Proceedings of the 2013 {Chilean} {Conference} on {Human} - {Computer} {Interaction}}}.\hskip 1em plus 0.5em minus 0.4em\relax Temuco Chile: ACM, Nov. 2013, pp. 36--41.

\bibitem{stefanidi_real-time_2022}
Z.~Stefanidi, G.~Margetis, S.~Ntoa, and G.~Papagiannakis, ``\BIBforeignlanguage{en}{Real-time adaptation of context-aware intelligent user interfaces, for enhanced situational awareness},'' \emph{\BIBforeignlanguage{en}{IEEE Access}}, vol.~10, pp. 23\,367--23\,393, 2022.

\bibitem{tadokoro_robocup-rescue_2000}
S.~Tadokoro, H.~Kitano, T.~Takahashi, I.~Noda, H.~Matsubara, A.~Shinjoh, T.~Koto, I.~Takeuchi, H.~Takahashi, F.~Matsuno, M.~Hatayama, J.~Nobe, and S.~Shimada, ``\BIBforeignlanguage{en}{The {RoboCup}-{Rescue} project: a robotic approach to the disaster mitigation problem},'' in \emph{\BIBforeignlanguage{en}{Proceedings 2000 {ICRA}. {Millennium} {Conference}. {IEEE} {International} {Conference} on {Robotics} and {Automation}. {Symposia} {Proceedings} ({Cat}. {No}.{00CH37065})}}, vol.~4.\hskip 1em plus 0.5em minus 0.4em\relax San Francisco, CA, USA: IEEE, 2000, pp. 4089--4094.

\bibitem{tiemann_celidon_2020}
J.~Tiemann, O.~Fuhr, and C.~Wietfeld, ``\BIBforeignlanguage{en}{{CELIDON}: {Supporting} {First} {Responders} through {3D} {AOA}-based {UWB} {Ad}-{Hoc} {Localization}},'' in \emph{\BIBforeignlanguage{en}{2020 16th {International} {Conference} on {Wireless} and {Mobile} {Computing}, {Networking} and {Communications} ({WiMob})}}.\hskip 1em plus 0.5em minus 0.4em\relax Thessaloniki, Greece: IEEE, Oct. 2020, pp. 20--25.

\bibitem{umlauft_communication_2016}
M.~Umlauft, C.~Raffelsberger, A.~Kercek, A.~Almer, T.~Schnabel, P.~Luley, and S.~Ladstaetter, ``\BIBforeignlanguage{en}{A communication and multi-sensor solution to support dynamic generation of a situational picture},'' in \emph{\BIBforeignlanguage{en}{2016 3rd {International} {Conference} on {Information} and {Communication} {Technologies} for {Disaster} {Management} ({ICT}-{DM})}}.\hskip 1em plus 0.5em minus 0.4em\relax Vienna, Austria: IEEE, Dec. 2016, pp. 1--7.

\bibitem{walker_mixed_2021}
M.~Walker, Z.~Chen, M.~Whitlock, D.~Blair, D.~A. Szafir, C.~Heckman, and D.~Szafir, ``\BIBforeignlanguage{en}{A {Mixed} {Reality} {Supervision} and {Telepresence} {Interface} for {Outdoor} {Field} {Robotics}},'' in \emph{\BIBforeignlanguage{en}{2021 {IEEE}/{RSJ} {International} {Conference} on {Intelligent} {Robots} and {Systems} ({IROS})}}.\hskip 1em plus 0.5em minus 0.4em\relax Prague, Czech Republic: IEEE, Sep. 2021, pp. 2345--2352.

\bibitem{wang_person--interest_2013}
X.~Wang, X.~Zhao, V.~Prakash, Z.~Gao, T.~Feng, O.~Gnawali, and W.~Shi, ``\BIBforeignlanguage{en}{Person-of-interest detection system using cloud-supported computerized-eyewear},'' in \emph{\BIBforeignlanguage{en}{2013 {IEEE} {International} {Conference} on {Technologies} for {Homeland} {Security} ({HST})}}.\hskip 1em plus 0.5em minus 0.4em\relax Waltham, MA, USA: IEEE, Nov. 2013, pp. 658--663.

\bibitem{wang_method_2023}
M.~Wang, H.~Ji, M.~Jia, Z.~Sun, J.~Gu, and H.~Ren, ``\BIBforeignlanguage{en}{Method and application of information sharing throughout the emergency rescue process based on {5G} and {AR} wearable devices},'' \emph{\BIBforeignlanguage{en}{Scientific Reports}}, vol.~13, no.~1, p. 6353, Apr. 2023.

\bibitem{weichelt_augmented_2018}
B.~Weichelt, A.~Yoder, C.~Bendixsen, M.~Pilz, G.~Minor, and M.~Keifer, ``\BIBforeignlanguage{en}{Augmented {Reality} {Farm} {MAPPER} {Development}: {Lessons} {Learned} from an {App} {Designed} to {Improve} {Rural} {Emergency} {Response}},'' \emph{\BIBforeignlanguage{en}{Journal of Agromedicine}}, vol.~23, no.~3, pp. 284--296, Jul. 2018.

\bibitem{whitlock_designing_2019}
M.~Whitlock, K.~Wu, and D.~Szafir, ``\BIBforeignlanguage{en}{Designing for {Mobile} and {Immersive} {Visual} {Analytics} in the {Field}},'' Aug. 2019, arXiv:1908.00680 [cs].

\bibitem{wilchek2025ajna}
M.~Wilchek, K.~Luther, and F.~A. Batarseh, ``Ajna: A wearable shared perception system for extreme sensemaking,'' \emph{ACM Transactions on Interactive Intelligent Systems}, vol.~15, no.~1, pp. 1--29, 2025.

\bibitem{zhang_exploring_2024}
K.~Zhang, B.~R. Cochran, R.~Chen, L.~Hartung, B.~Sprecher, R.~Tredinnick, K.~Ponto, S.~Banerjee, and Y.~Zhao, ``\BIBforeignlanguage{en}{Exploring the {Design} {Space} of {Optical} {See}-through {AR} {Head}-{Mounted} {Displays} to {Support} {First} {Responders} in the {Field}},'' in \emph{\BIBforeignlanguage{en}{Proceedings of the {CHI} {Conference} on {Human} {Factors} in {Computing} {Systems}}}.\hskip 1em plus 0.5em minus 0.4em\relax Honolulu HI USA: ACM, May 2024, pp. 1--19.

\bibitem{arif_comparative_2019}
S.~M.~U. Arif, P.~Mazumdar, and F.~Battisti, ``\BIBforeignlanguage{en}{A {Comparative} {Study} of {Rendering} {Devices} for {Safety}-{Critical} {Applications} in {Operative} {Control} {Rooms}},'' in \emph{\BIBforeignlanguage{en}{2019 11th {International} {Symposium} on {Image} and {Signal} {Processing} and {Analysis} ({ISPA})}}.\hskip 1em plus 0.5em minus 0.4em\relax Dubrovnik, Croatia: IEEE, Sep. 2019, pp. 282--287.

\bibitem{balfour_what_2013}
R.~E. Balfour and B.~P. Donnelly, ``\BIBforeignlanguage{en}{The what, why and how of achieving urban telepresence},'' in \emph{\BIBforeignlanguage{en}{2013 {IEEE} {Long} {Island} {Systems}, {Applications} and {Technology} {Conference} ({LISAT})}}.\hskip 1em plus 0.5em minus 0.4em\relax Farmingdale, NY, USA: IEEE, May 2013, pp. 1--6.

\bibitem{lugtenberg_magicbook_2023}
G.~Lugtenberg, K.~Mori, Y.~Matoba, T.~Teo, and M.~Billinghurst, ``\BIBforeignlanguage{en}{The {MagicBook} revisited},'' in \emph{\BIBforeignlanguage{en}{2023 {IEEE} {International} {Symposium} on {Mixed} and {Augmented} {Reality} {Adjunct} ({ISMAR}-{Adjunct})}}.\hskip 1em plus 0.5em minus 0.4em\relax Sydney, Australia: IEEE, Oct. 2023, pp. 801--806.

\bibitem{nilsson_using_2009}
S.~Nilsson, B.~Johansson, and A.~Jonsson, ``\BIBforeignlanguage{en}{Using {AR} to support cross-organisational collaboration in dynamic tasks},'' in \emph{\BIBforeignlanguage{en}{2009 8th {IEEE} {International} {Symposium} on {Mixed} and {Augmented} {Reality}}}.\hskip 1em plus 0.5em minus 0.4em\relax Orlando, FL, USA: IEEE, Oct. 2009, pp. 3--12.

\bibitem{peretti_augmented_2022}
O.~Peretti, Y.~Spyridis, A.~Sesis, G.~Efstathopoulos, A.~Lytos, T.~Lagkas, and P.~Sarigiannidis, ``\BIBforeignlanguage{en}{Augmented reality training, command and control framework for first responders},'' in \emph{\BIBforeignlanguage{en}{2022 7th {South}-{East} {Europe} {Design} {Automation}, {Computer} {Engineering}, {Computer} {Networks} and {Social} {Media} {Conference} ({SEEDA}-{CECNSM})}}.\hskip 1em plus 0.5em minus 0.4em\relax Ioannina, Greece: IEEE, Sep. 2022, pp. 1--5.

\bibitem{stone_mixed_2017}
R.~J. Stone, R.~Guest, P.~Mahoney, D.~Lamb, and C.~Gibson, ``\BIBforeignlanguage{en}{A ‘mixed reality’ simulator concept for future {Medical} {Emergency} {Response} {Team} training},'' \emph{\BIBforeignlanguage{en}{Journal of the Royal Army Medical Corps}}, vol. 163, no.~4, pp. 280--287, Aug. 2017.

\bibitem{doswell2020juxtopia}
J.~T. Doswell, J.~Jolmson, B.~Brockingon, A.~Mosby, S.~Salaam, and A.~Chinery, ``Juxtopia{\textregistered} cammrad prepare: A wearable ai-ar platform for clinical training emergency first response teams,'' in \emph{2020 22nd Symposium on Virtual and Augmented Reality (SVR)}.\hskip 1em plus 0.5em minus 0.4em\relax IEEE, 2020, pp. 164--168.

\bibitem{friedman2024prehospital}
N.~Friedman, M.~Zuniga-Hernandez, J.~Titzler, M.~Y. Suen, E.~Wang, O.~Rosales, J.~Graham, P.~D’Souza, M.~Menendez, and T.~J. Caruso, ``Prehospital pediatric emergency training using augmented reality simulation: A prospective, mixed methods study,'' \emph{Prehospital Emergency Care}, vol.~28, no.~2, pp. 271--281, 2024.

\bibitem{koutitas2019virtual}
G.~Koutitas, K.~S. Smith, G.~Lawrence, V.~Metsis, C.~Stamper, M.~Trahan, and T.~Lehr, ``A virtual and augmented reality platform for the training of first responders of the ambulance bus,'' in \emph{Proceedings of the 12th ACM International Conference on PErvasive Technologies Related to Assistive Environments}, 2019, pp. 299--302.

\bibitem{majumdar_cloud-based_2023}
S.~Majumdar and S.~Kirkley, ``\BIBforeignlanguage{en}{A cloud-based fire safety system for emergency responders and civic community},'' in \emph{\BIBforeignlanguage{en}{2023 {IEEE} {Long} {Island} {Systems}, {Applications} and {Technology} {Conference} ({LISAT})}}.\hskip 1em plus 0.5em minus 0.4em\relax Old Westbury, NY, USA: IEEE, May 2023, pp. 1--6.

\bibitem{endsley1995toward}
M.~R. Endsley, ``Toward a theory of situation awareness in dynamic systems,'' \emph{Human Factors}, vol.~37, no.~1, pp. 32--64, 1995.

\bibitem{smets_effects_2008}
N.~J. J.~M. Smets, G.~M. Te~Brake, M.~A. Neerincx, and J.~Lindenberg, ``\BIBforeignlanguage{en}{Effects of mobile map orientation and tactile feedback on navigation speed and situation awareness},'' in \emph{\BIBforeignlanguage{en}{Proceedings of the 10th international conference on {Human} computer interaction with mobile devices and services}}.\hskip 1em plus 0.5em minus 0.4em\relax Amsterdam The Netherlands: ACM, Sep. 2008, pp. 73--80.

\bibitem{wilson_head-mounted_2009}
J.~Wilson and P.~Wright, ``\BIBforeignlanguage{en}{Head-mounted display efficacy study to aid first responder indoor navigation},'' \emph{\BIBforeignlanguage{en}{Proceedings of the Institution of Mechanical Engineers, Part C: Journal of Mechanical Engineering Science}}, vol. 223, no.~3, pp. 675--688, Mar. 2009.

\end{thebibliography}
